\documentclass[12pt]{article}
\usepackage{amssymb}
\usepackage{amsmath}
\usepackage{graphicx}

\newcommand{\be}{\begin{equation}}

\newcommand{\ee}{\end{equation}}

   \def\(({\left(}
   \def\)){\right)}
\def\[[{\left[}

\def\]]{\right]}

\def \form#1 {eq. (\ref{#1}) }
\def \parziale#1#2  {{\partial {#1} \over \partial {#2}}}

\newcommand{\ba}{\begin{eqnarray}}
\newcommand{\ea}{\end{eqnarray}}

\newcommand{\al}{\alpha}

\newcommand{\e}{\varepsilon}

\begin{document}

\title{{\sc Geometry of the Casimir effect}\footnote{
{\it Proceedings of the 15 th SIGRAV Conference on General Relativity and Gravitational Physics},  
Villa Mondragone, Monte Porzio Catone, Roma, Italy, September 9-12, 2002; to appear in 
{\it Recent Developments in Gravitational Physics}, Institute of Physics Conference Series 176, Ed. Ciufiolini et al, Oct 2004}}

\author{Roger {\sc Balian}\footnote{E-mail: balian@spht.saclay.cea.fr}  and Bertrand {\sc Duplantier}\footnote
{E-mail: bertrand@spht.saclay.cea.fr}\\
Service de Physique Th{\'e}orique\footnote{Unit{\'e} de recherche associ{\'e}e au CNRS 2306}\\
Orme des Merisiers,  
CEA-Saclay\\ 
F-91191 Gif-sur-Yvette Cedex}

\date{}
\maketitle

\begin{abstract}

When the vacuum is partitioned by material boundaries with
  arbitrary shape, one can define the zero-point energy and the free
  energy of the electromagnetic waves in it: this can be done, independently of the
  nature of the boundaries, in the limit that they become perfect
  conductors, provided their curvature is finite. The first examples we consider are Casimir's original configuration of
   parallel plates, and the experimental situation of a sphere in front of a plate. For arbitrary geometries, we give an explicit
  expression for the zero-point energy and the free energy in
  terms of an integral kernel acting on the boundaries; it can be
  expanded in a convergent series interpreted as a succession of an
  even number of scatterings of a wave. The quantum and thermal
  fluctuations of vacuum then appear as a purely geometric
  property. The Casimir effect thus defined exists only owing to the
  electromagnetic nature of the field. It does not exist for thin
  foils with sharp folds, but Casimir forces between solid wedges are
  finite.  We work out various applications: low temperature, high temperature where wrinkling constraints
  appear, stability of a plane foil, transfer of energy from one side
  of a curved boundary to the other, forces between distant
  conductors, special shapes (parallel plates, sphere, cylinder,
  honeycomb).
\end{abstract}

\section{Introduction}
\subsection{A brief history}
According to Planck (1900) \cite{planck}, the energy of a stationary mode of the electromagnetic field (with
frequency $\omega/2\pi$)
 is quantized  with a contribution
$\hbar\omega$ per photon. Because of the quadratic structure of the electromagnetic energy, a canonical transformation
describes each mode as an harmonic oscillator. At the classical level, this representation was already known to Rayleigh, and  Planck
guessed rightly at the quantization of the oscillator energy levels. It was later discovered
that their complete form reads
\begin{equation}
\e_N=\hbar \omega\, (N+\frac{1}{2}),\ N \in {\bf N},
\label{1/2}
\end{equation}
with $N$ the number of photons, and with a non-vanishing zero-point energy, $\frac{1}{2}\hbar
\omega$, which reflects Heisenberg's uncertainty principle. One might have believed that
the reference vacuum energy would be unobservable. Hendrik B G Casimir showed to the contrary in his famous 1948 article
\cite{casimir}  that a physical force can be generated by vacuum fluctuations (see also \cite{caspol}, and 
for an historical account, \cite{SarSpar}).

In its original form, the Casimir effect describes an attraction
between two plane, parallel, perfectly conducting plates, which is
explained by this occurence of a virtual electromagnetic field in the
vacuum separating the plates and in the vacuum lying outside them. Indeed, even
 when no real photon is present, the
 zero-point motion of each mode $m$ of the field  yields a contribution $\frac 1 2
\hbar \omega_m(L)$ to the energy, which depends on the distance $L$
between the plates. The variation with $L$ of the overall zero-point
energy manifests itself as the Casimir force. It is remarkable that,
in spite of the divergence which appears when summing the zero-point
energy over all modes, one finds a finite value
for this force. Let us remark that a dimensional argument allows one to anticipate the right form of Casimir's result.
A quantum pressure between two large plates should indeed depend on $\hbar$, $c$ and on the distance $L$ between them.
To get pressure units  $P \propto {\rm N}\ {\rm
m}^{-2}= {\rm J}\ {\rm m}^{-3}$, with $\hbar \propto {\rm J\ s}$, $c \propto
{\rm m}\ {\rm s}^{-1}$, and $L\propto {\rm m}$, one has first to multiply
$\hbar$ by $c$, so as to eliminate time, and then divide by $L^{-4}$. Hence $\displaystyle  P_{\rm Casimir} \propto {
\frac{\hbar c}  {L^4}}$, and the real surprise is the finiteness of the numerical coefficient!

In recent experiments conducted in 1997-98 (\cite{lam,moh}), the Casimir force between
a metallized sph{e}re and a metallized  plate was finally measured definitively with the help of atomic force microscopes.
The field is nowadays the subject of intense research activity \cite{bordag,bourbaphy}.  
Measurements of attractive Casimir forces are performed for the sphere-plane geometry \cite{harris}, 
 for micromechanical torsional devices \cite{chan}, and for the original Casimir's planar geometry \cite{carugno}. 
Lateral Casimir forces are measured between corrugated surfaces \cite{chen1,chen2} 
(see also \cite{emig}). 
Roughness effects are experimentally important (\cite{genet} and references therein). The experimental detection of temperature effects is an open problem \cite{chen3}. 

\subsection{Statement of the problem}
It is somewhat puzzling to regard the Casimir effect as a property of
vacuum containing a virtual electromagnetic field. It may look more natural
to attribute it to the matter of the plates. Actually it is possible
through Maxwell's equations to express the electromagnetic field, and
hence its energy, in terms of the charge and current densities of the
particles which can move within the conducting plates and which are
the sources for the field. The zero-point motion of these particles
then yields a non-vanishing interaction energy although the
expectation values of the charge and current densities vanish. The
force between the conducting plates at zero temperature can then be
interpreted as a result of the interaction between the virtual
zero-point currents that must exist in the ground state of matter owing to
Heisenberg's inequality. However, this viewpoint is plagued by the
fact that the interaction between charged particles is not
instantaneous, but retarded. On the other hand, the evaluation of the
force as a result of this interaction would rely on the specific
structure of matter. In contrast, Casimir's viewpoint, which focuses
on the field rather than on the matter of the plates, shows that the
effect is nearly independent of the properties of matter (provided the
plates are good conductors); moreover, as function of the
electromagnetic field ${\bf E, B}$ the energy of this field at a given time is
simply expressed as
\begin{equation}\label{1}
\mathcal{E}_{\rm f}= \frac 1 2 \int d^3r [\epsilon_0 {\bf E}^2(r) +
\mu^{-1}_0 {\bf B}^2(r)]
\end{equation}
in terms of the field at the same time, whereas ${\bf E}(r)$ and ${\bf
  B}(r)$ depend on the charges and currents at earlier times.

We wish to study the energy $\mathcal{E}_{\rm f}$ of the
electromagnetic field in empty regions of space limited by boundaries
with arbitrary shape, under circumstances when the expectation value
$\langle{\bf E}\rangle,\langle{\bf B}\rangle$ of the quantum field vanishes at any point. The
energy (\ref{1}) may be non-zero for two reasons.

On the one hand, the uncertainty relations prevent the {\it quantum
  fluctuations} of ${\bf E}$ and ${\bf B}$ from vanishing, since these
  operators do not commute; hence the expression (\ref{1}) for the
  energy has a positive minimum, the zero-point energy of the
  field. (We shall see that this value is not only strictly positive, but in fact
  infinite; however its variations are finite and physically
  meaningful.)

On the other hand, the vacuum where the field is considered is bounded
by walls with which the field can be in thermodynamic equilibrium, at
some temperature $T$. The field therefore presents random {\it thermal
  fluctuations} around its vanishing expectation value; they
contribute to the energy (\ref{1}) and to the entropy. We shall study
this problem by evaluating the free energy of the field, wherefrom
all thermodynamic equilibrium properties follow; in particular its
variations with the shape provide the
constraints on the walls at fixed temperature. This free energy includes the zero-point
energy, to which it reduces at $T=0$. We shall thus treat
simultaneously the {\it Casimir effect} proper, associated with the
{\it zero-point} energy, that is, to virtual photons, and the {\it
  radiation pressure} effects for a {\it black-body} with arbitrary
shape, which are associated with real photons in equilibrium with the
walls that act as a thermal bath.

Since general relativity is of no relevance in the present problem,
energy is defined within an additive constant. We are interested only
in its variations and can thus get rid of divergences in the theory by
substracting some constant that will tend to infinity.

We wish to define the Casimir effect as a property belonging only to
the {\it field in vacuum}, independently of the nature of the material in the
walls. In general the zero-point energy of a field depends on the
matter to which this field is coupled. We shall use the term ``Casimir
effect'', in contrast to some authors, only when this energy can be
defined separately. For electromagnetic fields within real materials,
the interaction between the field and
the charges does not allow us in general to separate out the energy
(\ref{1}) of the field alone; moreover the presence of a material
affects the field even outside it. However, a complete decoupling is
achieved in the limit of {\it perfectly conducting} boundaries, which
can be approached experimentally by use of {\it superconductors}. Both
the electric and the magnetic fields ${\bf E}$ and ${\bf B}$ vanish
inside them. Outside them, they simply impose the boundary conditions
\begin{equation}\label{2}
{\bf E}_{\rm t} = 0\ \ ,\ \ {\bf B}_{\rm n} = 0
\end{equation}
on the tangential ${\rm (t)}$ and normal ${\rm (n)}$ components of the
field. The presence of material bodies then has only a mere {\it
  geometric effect}. The part $-\int{\bf j}.{\bf A}$ of the energy,
 which involves the currents and the vector potential and which determines the
matter-radiation coupling in the equations of motion, can be assigned to the
matter and left aside while ${\mathcal E}_{\rm f}$ is assigned to the
vacuum. Anyhow, for perfect conductors, in a gauge where ${\bf E} =
-\partial{\bf A}/\partial t$, this coupling energy vanishes on average since
${\bf A}$ is perpendicular to the surface current ${\bf j}$.

In this idealized model, the field and the matter of the boundaries do
not exchange any energy, even in time-dependent situations, although
the coupling between the potentials and the charged particles relates
$\bf E$ and $\bf B$ to the charge and current densities through
Maxwell's equations of motion. Indeed, the rate of decrease of the energy
(\ref{1}) in some empty region of space is the outgoing flux of the
Poynting vector $\mu^{-1}_0 {\bf E} \times {\bf B}$ across the
boundary of this region. The conditions (\ref{2}) imply that the
Poynting vector is tangent to a perfectly conducting wall, and hence
that no energy can flow across such a wall. The establishment of
thermal equilibrium in a vacuum is ensured only by the fact that real
conductors are never perfect; this allows energy transfers between
field and matter. Similarly to the model of an ideal gas, the present model is too
crude to describe the establishment of equilibrium, but it is adequate
for equilibrium properties.

\subsection{Synopsis}
We first recall Casimir's calculation for parallel plates, as well as its generalization to arbitrary temperatures.
Through the further use of the so-called Derjaguin approximation, this allows us to briefly describe a
recent experiment, and the irrelevance
of temperature effects for the latter, allowing the conclusion that macroscopic quantum vacuum fluctuations
are indeed observed at room temperature!

In the further study of arbitrary conductor geometries, we shall exhibit general aspects of the
Casimir effect, for walls with arbitrary shape and at arbitrary
temperatures. We wish to answer a few theoretical questions. Can the
energy associated with the quantized electromagnetic field in the
empty regions bounded by conducting walls be defined independently of
the properties of these walls? How does it vary with the temperature
of the photon gas that constitutes the field? Does the existence of
the Casimir effect depend on the specific features of
electromagnetism?

We shall rely on two detailed articles \cite{BD1,BD2} which deal with the above
questions. This will allow us to leave aside the technicalities and to
focus on the various ideas.  The main result is embedded in section \ref{sec.renormalized}. For a bibliography on the
Casimir effect we refer the reader to articles published in \cite{bourbaphy}, and to recent
monographs \cite{bordag,levin,milonni,most,eliz,krech,brank,milton}.

\section{Casimir's calculation (1948)}
\label{sec.casimir}
Consider two identical plates, parallel to the plane $yOz$, with a large area ${\cal A}=L_y\times L_z$, and
separated by a
distance $L$ from each other along the orthogonal direction $Ox$. \par
\begin{figure}
\begin{center}
\includegraphics[angle=0,width=.55\linewidth]{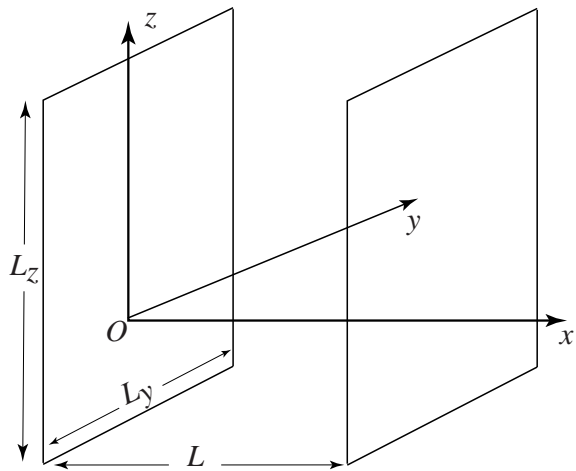}
\end{center}
\caption{{Conducting plates configuration.}}
\label{plates}
\end{figure}
The ideal plates are perfectly conducting, and the stationary modes are described by  wave vectors
$ (k_x, {\bf k}),$ where ${\bf k}=(k_y, k_z)$ is parallel to the plates ; in the perpendicular direction,
the boundary conditions give the discrete series: $\displaystyle k_x = \frac{\pi n}{L},$ where $n \in \mathbb N$, while
periodic boundary conditions along the plates\footnote{As usual, the asymptotic result does not depend on these
boundary conditions.} give $\displaystyle  (k_y,
k_z)=\left(\frac{2\pi n_y}{L_y},\frac{2\pi n_z}{L_z}\right),$ where
 $n_y,n_z \in {\mathbb Z}.$ This eigenmode, denoted by $(n,{\bf k}),$ oscillates with a ``frequency'' $\omega$
\begin{equation} \omega=\omega_n({{\bf k}}) = c
\sqrt{\frac{\pi^2 n^2}{L^2}+{\bf k}^2}.
\label{om}
\end{equation}
Each eigenmode occurs with two possible polarizations  (except for $n=0$).

{At} zero temp{e}rature, the cavity electromagnetic energy is the sum
 ${\cal E}_0$ of the zero-point energies of the eigenmodes
\begin{equation}
\label{pz}
{\cal E}_0=\sum \limits_{{\rm modes}\ (n,{\bf k})}\varepsilon_0[\omega_n({\bf
k})]\ ,\ \ \varepsilon_0(\omega)=\frac{1}{2}\hbar \omega.
\end{equation}
The terms of this series are unbounded, and the series diverges. However, physics tells us how to cure this problem: at very high frequencies, the material ceases
to conduct and becomes a dielectric, ultimately transparent to radiation. Then the boundary
conditions no longer apply, and very high eigenmodes no longer contribute to the resulting force.
This leads to introduce the mathematically convenient regularization
\begin{equation}
{\cal E}_0=\sum \limits_{{\rm
modes}\ (n,{\bf k})}\varepsilon_0[\omega_n({\bf k})] \
\chi\left(\frac{\omega_n({\bf k})}{\omega_c}\right),
\label{e0}
\end{equation}
where the cut-off function $\chi(\omega/\omega_c)$, is such that
$\chi(0)=1$, and is  regular at the origin. It vanishes, along with all its derivatives,
for $\omega/\omega_c\to +\infty$, sufficiently fast so that the sum converges.
The cut-off frequency $\omega_c$ appears in $\chi$ for dimensional reasons ; it depends on the microscopic
characteristics of the material.
(The
{\it perfect} conductor limit corresponds to $\omega_c \to +\infty$, so that
$\chi({\omega / \omega_c}) \to 1$ for any finite $\omega$.)\par
\hskip 2mm For large values of $L_y,\,L_z$, one can replace the sums over
{\it parall{e}l} wave vectors by integrals. The distance $L$ remains finite and associated with discrete modes. On thus gets
\begin{equation}  \sum _{{\rm modes}\ (n,{\bf k})}\cdots=2 \frac{{\cal A}}{ (2
\pi)^2}\sum^{\infty}_{n=0}{}^\prime  \int_{{\bf R}^2} {\rm
d}^2{\bf k}\cdots, \label{modes}
\end{equation}
where the {\it prime}  means that the  $n=0$ mode has weight $1/2$. Let us define
$\varepsilon(\omega)=\varepsilon_0(\omega) \
\chi\left(\omega/\omega_c\right).$ The energy (\ref{e0}) can then be written from (\ref{modes})
$${\cal E}_{0}(L)=2 \frac{{\cal A}}{ (2
\pi)^2}\sum _{n=0}^{\infty}{}^\prime  \int_{{\bf R}^2} {\rm
d}^2{\bf k} \ \varepsilon \left[{\omega}_n({\bf k})\right].$$
Owing to (\ref{om}) on has, for $n$ fix{ed}, the form $\omega {\rm
d}\omega =c^2 k {\rm d} k$, where $k=|{\bf k}|$.
By simple integration over parallel wave vectors ${\bf k}$:
\begin{equation}
\label{int}
\int_{{\bf R}^2} {\rm d}^2{{\bf k}}\,
\varepsilon \left[\omega_n({\bf k})\right] =
 2 \pi
c^{-2}\int_{\omega_n({\bf 0})}^{+\infty} \omega\,{\rm d}{\omega}\ \varepsilon(\omega), \ \ {\omega_n({\bf 0})}=\pi c n/L,
\end{equation}
whence:
\begin{equation}
  {\cal E}_0 = {\cal A}
\frac{1} {\pi c^2} \sum \limits _{
n=0}^{\infty}{}^\prime\int\limits^{\infty}_{\pi c n/L}  {\rm d}\omega
\ \omega \varepsilon_0(\omega)
\chi\left(\frac{\omega} {\omega_c}\right).
\end{equation}
The associated force, $X_0$, can be readily derived,
\begin{equation}
X_0=-\frac{\partial  {\cal E}_0}
{\partial L}=-{\cal A} \frac{\pi^2 \hbar c} {2L^4}
\sum \limits_{ n=0}^{\infty}{}^\prime g(n),
\hskip.2cm  g(n)=n^3 \chi\left(\frac{\pi c n} {L \omega_c}\right).
\label{X_0}
\end{equation}

The {e}quivalent ${X}_0^{\infty}$ of ${X}_0$  in the large $L$ limit
is given, as any continuum limit, by substituting the integral over $n$ for
the ``primed'' sum over $n$:
\begin{equation}
\label{X_0inf}
X_0^{\infty}=-{\cal
A} \frac{\pi^2 \hbar c} {2L^4}\int \limits^{\infty}_{0} {\rm d}n
\ g(n).
\end{equation}

To obtain the zero-temperature force, associated with the vacuum energy
${\varepsilon}_0(\omega)=\frac 1 2\hbar \omega $,
and acting on the plate, one must also take into account the opposite force exerted by the (infinite)
electromagnetic vacuum {\it outside} of the capacitor. This force is just the opposite of
 (\ref{X_0inf}), whence the resulting force
$${\tilde X}_0=X_0-X_0^{\infty}=-{\cal
A} \frac{\pi^2 \hbar c} {2L^4} \left[    \hbox{ $   \sum \limits_{
n=0}^{\infty}{}^\prime g(n) - \int \limits^{\infty}_{0} {\rm d}n \,
g(n)$}\right].$$
To evaluate  the
 difference between a series and the associated integral, we use the  {\it
Euler-Maclaurin} formula:
\begin{equation}\hbox{ $ \sum \limits_{ n=0}^{\infty}{}^\prime
  g(n) - \int \limits^{+\infty}_{0} dn\ g(n)$}= -\frac 1 {12} g'(0)+
  \frac 1
  {6!} g'''(0)+ \mathcal O\left( g^{[5]}(0)\right),\end{equation}
which involves all $g$'s derivatives of odd order, taken at the origin, and which is valid for a function
 $g$ vansishing at infinity, as well as all its derivatives.  By calculating the successive derivatives, one finds here
$$g'(0)=0, \ g'''(0)=6\chi (0)=6,\
g^{[p]}(0)=\mathcal O\left({\omega_c}^{-(p-3)}\right), \ p \ge 3.$$
One therefore finds the finite value\footnote{The presence of the factor $n^3$,
 varying rapidly with $n$, yields the non-vanishing value $g'''(0)=3!$. In the absence of such a term,
 the Euler-Maclaurin formula would start with $g'(0)=\mathcal O\left(
{\omega_c}^{-1}\right)$, and the diff{e}rence between the sum and the integral
would  vanish in the perfect conductor limit.} $$ \hbox{ $ \sum \limits_{ n=0}^{\infty}{}^\prime
  g(n) - \int \limits^{\infty}_{0} dn g(n)$}=  \frac 1
{5!} + \mathcal O\left( {\omega_c}^{-2}\right).$$
The zero-temperature resulting force thus possesses a universal limit for perfect conductors,
i.e, when $\omega_c \to +\infty$. The limit pressure,
found by  H. B. G.  Casimir in 1948, is:
\begin{equation}\frac 1  {\cal A}{\tilde
X}_0=-\frac{{\pi}^2} {240} \frac{ \hbar c} {{L}^4}.\label{Casimir}\end{equation}
The Casimir force is {\it attractive}, and one finds the analytic form which was anticipated in terms of
$\hbar$, $c$ and of the length $L$. Only the numerical coefficient remained to be found: $-\pi^2/240$, and the
remarkable fact is that it is non-vanishing and
{\it universal}, i.e., ind{e}pendent of the microscopic nature of the perfect conductors.
To the resulting Casimir force is associated a {\it subtracted} zero-point energy $\tilde {\cal E}_0$ such that
\begin{equation}
\label{energiecasimir}
\frac{1}{\cal A}\tilde {\cal E}_{0}=- \frac{\pi^2} {720}\frac{\hbar c} {L^3},\ \
\tilde X_0=-\frac{\partial \tilde {\cal E}_0}{\partial L}.
\end{equation}
\section{Electromagnetic free energy of a planar capacitor}
\subsection{Vacuum and thermal parts}
Let us briefly consider the effect of temperature, in order 
to compare it to the 
zero-point effect\footnote{The first calculations are due to Fierz \cite{fierz} and Mehra \cite{mehra}.}.
At a given temperature  $T$,  photons
will fill in the cavity, and they obey the ``black-body'' statistics.
%
To each classical eigenmode $(n,{\bf k})$ is
 associated the Hamiltonian of a quantum harmonic oscillator, with frequency $\nu=\omega/2\pi$, $\omega=\omega_n({\bf k})$,
 given by (\ref{om}).
The eigenvalues of this Hamiltonian are then
$\e_N=\hbar \omega (N+1/2)$, where $\ N \geq 0$ is the number of photons in the mode.

The free energy of an eigenmode $\omega$ at temperature $T$, with $\beta=1/k_B T$, and $k_B$
{Boltzmann's constant}, has the form
\begin{equation}
f(\omega)=-\frac{1}{\beta} {\rm ln} \sum_{N=0}^{\infty} e^{-\beta
  \hbar \omega (N+\frac{1}{2})}=\frac 1 2 \hbar \omega+
\frac 1 {\beta}{\rm ln}
\left(1-e^{-\beta \hbar \omega}\right).
\label{freeenergy}
\end{equation}
We recast it as
\begin{equation}
f(\omega)=\varepsilon_0(\omega)+f_{\rm T}(\omega),
\end{equation}
with
\begin{equation}
\varepsilon_0(\omega)=\frac 1 2 \hbar \omega, \hskip.1cm f_{\rm
T}(\omega)={\beta}^{-1}\varphi\left({\beta \hbar
\omega}\right),
\end{equation}
where  $f_{\rm T}(\omega)$ is the {\it
thermal} part of the mode free energy,  and where
$$\varphi(x)={\rm ln} \left(1-e^{-x}\right),\ \ \varphi(x)\le 0.$$
By definition, the zero-temperature limit of $f_{\rm T}(\omega)$ vanishes.

The purely thermal part of the electromagnetic free energy between the plates is then defined as follows:
\begin{equation}
\label{F}{\cal F}_{\rm T}= \sum_{(n,{\bf k})\;{\rm modes}}
f_{\rm T}[\omega_n({\bf k})] .
\end{equation}
In contrast with the eigenmode sum associated with the vacuum, the sum
 (\ref{F}) associated with the thermal radiation is {\it convergent}.
By using (\ref{modes}) it can be written as
\begin{equation}
\label{F*}{\cal F}_{\rm T}(L)
=2 \frac{{\cal A}}{ (2
\pi)^2}\sum _{n=0}^{\infty}{}^{\prime}  \int_{{\bf R}^2} {\rm
d}^2{\bf k} \ {\beta}^{-1}\varphi\left[{\beta \hbar
\omega}_n({\bf k})\right].
\end{equation}
 In
(\ref{F*}) the integration over parallel vectors
${\bf k}$ gives, as in (\ref{int}):
\begin{equation} \label{int'} \int_{{\bf R}^2} {\rm d}^2{\bf
k}\,\varphi[{\beta \hbar
\omega_n({\bf k})}]=2 \pi c^{-2} \int_{\omega_n({\bf 0})}^{+\infty} \omega\,{\rm
d}{\omega}\ \varphi(\beta \hbar
\omega),
\end{equation}
with ${\omega_n({\bf 0})}=\pi nc/L$.
In terms of the dimensionless variable $x=\beta \hbar
\omega$, and of
\begin{equation}
\label{psi}
\psi (u)=\int \limits ^{ +\infty}_ {u } {\rm d}x \ x
\varphi(x),
\end{equation}
the free energy is obtained as a simple series:
\begin{equation}
\label{FS}{\cal F}_{\rm T}(L)=2 \frac{\cal A}
{2\pi \beta} \frac 1  {(\beta \hbar c)^2} \sum _{n=0}^{\infty}{}^{\prime}
\psi (n\alpha),\,\,\,\al =\beta \pi \hbar c /L.
\end{equation}
\subsection{Continuous limit}
By comparing the energies $\hbar \omega$ of
photons belonging to two consecutive eigenmodes, on can estimate
the domain of temperatures $T$ or separations $L$ for which the discrete character
of the eigenmodes disappears. For vanishing parallel wave vectors: $\hbar\Delta
\omega=\hbar[\omega_{n+1}({\bf 0})-\omega_n({\bf 0})]=\hbar\pi
c/L={\al} k_B T$, and  the eigenmodes appear as a continuum for $\al \leq 1$. At ordinary temp{e}rature, $T \simeq 300\, {\rm K}$, this gives
$L \geq 24\,\mu {\rm m}$, and only for shorter distances will the discrete character of the eigenmodes be detectable.
\par
Let us introduce the large $L$ limit, ${\cal F}_{\rm T}^{\infty}(L)$, of the free energy
${\cal F}_{\rm T}(L)$ (\ref{FS}). In the limit  $\alpha=\beta \pi \hbar c/L \ll
1$, the series in ${\cal F}_{\rm T}$
converges towards the integral{\footnote{The presence of the ``prime'' notation in the sum over $n$ and of the factor $\frac{1}{2}$ for the $n=0$ mode
are irrelevant in the large
 $L$ limit. They imply a finite difference between the sums, which are here evaluated at order $\mathcal O(L)$ as integrals.}}\par
$${\cal F}_{\rm T}^{\infty}
=2 \frac{\cal A} {2\pi \beta}\frac 1 {(\beta \hbar c)^2} \frac{1}{\al}
\int \limits_{0}^{ +\infty} {\rm d}{u}\ \psi (u).$$
The numerical coefficient is
$$\int \limits_{0}^{ +\infty} {\rm
d}{u}\ \psi (u)=\int \limits ^{ +\infty}_ {0 } {\rm d}x \ x^2 \varphi(x)
=-2 \zeta (4)= -\frac{{\pi}^4} {45},$$
in terms of the Riemann $\zeta$ function.
The continuum free energy
\begin{equation}
\label{Fnoir}
{\cal F}_{\rm T}^{\infty}=- 2 \frac{\cal A}
{2\pi \beta}\frac 1 {(\beta \hbar c)^2}\frac 2 {\alpha}\zeta
(4)=-\frac{{\cal A} \ L} {\beta}\frac 1 {(\beta \hbar c)^3}\frac{{\pi}^2}{45},
\end{equation}
is precisely the {\it black-body free energy} in a large volume $\Omega={\cal A} \times L$.\par

It is convenient to introduce the thermal {\it subtracted} free energy
\begin{equation} \label{Ftilde}
{\tilde {\cal F}}_{\rm T}={\cal F}_{\rm T}- {\cal F}_{\rm T}^{\infty},
\end{equation}
which reads, owing to (\ref{FS}) and (\ref{Fnoir})
\begin{eqnarray}
\label{Ftilde'}{\tilde {\cal F}}_{\rm T}
=2\frac{\cal A}
{2\pi \beta}\frac 1 {(\beta \hbar c)^2} \left[\sum _{n=0}^{\infty}{}^{\prime}
\psi (\al n)+\frac 2 {\alpha} \zeta
(4)\right].
\end{eqnarray}
\subsection{Thermal forces}
The force on, e.g., the right hand plate, is calculated as
\begin{equation} \label{Xdef}
X_{\rm T}(L)=-\frac{\partial {\cal F}_{\rm T}(L)} {\partial L}.
\end{equation}
From the free energy (\ref{FS}) one finds
\begin{equation} \label{X} X_{\rm T}=-2 \frac{\cal A}
{2\pi \beta L}\frac 1{(\beta \hbar c)^2} \sum \limits_{ n=0}^{ \infty}
{}^{\prime} n^2\al^2 \varphi (n\al).
\end{equation}
This force is perpendicular to the plate and positive ($\varphi(x)\leq 0),$ hence r{e}pulsive.
It is the black-body pressure in a finite geometry.
In the  $L\to \infty$ limit, one recovers the infinite volume black-body radiation pressure \begin{equation} \label{Xi}
X_{\rm T}^{\infty}=
-\frac{\partial {\cal F}_{\rm T}^{\infty}}{\partial L}
=\frac{\cal A} {\beta}\frac 1 {(\beta \hbar c)^3}\frac{{\pi}^2}
{45}.
\end{equation}
In fact, a given plate at temperature $T$ is in equilibrium with the thermal radiation existing
on both sides. It thus will also be subjected to the pressure force exerted by the exterior photons, which is just the
infinite volume black-body pressure force, $-X_{\rm T}^{\infty}$, that we just evaluated. The resulting thermal force is therefore:
\begin{equation} \label{Xtilde}
{\tilde X}_{\rm T}=X_{\rm T} -X_{\rm T}^{\infty}=-\frac{\partial
\left({\cal F}_{\rm T}-{\cal F}_{\rm T}^{\infty} \right)}{\partial L}
=-\frac{\partial {\tilde {\cal F}}_{\rm T}}{\partial L}. 
\end{equation}
\subsection{Short distance or low temperature expansion}
The  series (\ref {X}) giving the interior thermal force ${X}_{\rm T}$ yields a natural {\it low-temperature}
 or  {\it short-distance} expansion
for $\al=\beta \pi \hbar c /L \gg 1$. Indeed the $n=0$ term vanishes, and for $n\geq 1$,
$\varphi(n\al) \sim -e^{-n\al}$.
The $n=1$ mode thus contributes a leading exponentially small r{e}pulsive force:
$$X_{\rm T}={\cal A}\frac 1
{\beta} \frac{\pi} {L^3}\left[e^{-\alpha} +\mathcal O\left(e^{-2\al}\right)\right],\ \al
\gg 1.$$
In the thermal
{\it r{e}sulting force} (\ref{Xtilde}) ${\tilde X}_{\rm T}$, the leading term will thus be the black-body
exterior term:
\begin{equation}
\label{tildeXT}
\frac {1} {\cal A}{\tilde X}_{\rm T}=-\frac{\pi^2}{45}\frac 1
{\beta}\frac 1 {(\beta \hbar c)^3}+ \frac 1
{\beta} \frac{\pi} {L^3}\left[e^{-\alpha} +\mathcal O\left(e^{-2\al}\right)\right].
\end{equation}
\subsection{Comparison of zero-point and thermal effects}
Adding (\ref{Casimir}) and (\ref{tildeXT}), one gets the total pressure at low temperature or short distance:
\begin{equation}
\label{Xtot}
\frac 1 {\cal A}{X}
=\frac 1 {\cal A}\left({\tilde X}_0+\tilde X_{\rm
    T}\right)=-\frac{\pi^2} {240} \frac{\hbar c} {{L}^4}
-\frac{\pi^2}{45}\frac 1
{\beta}\frac 1 {(\beta \hbar c)^3}+ \frac 1
{\beta} \frac{\pi} {L^3}e^{-\alpha}+\cdots.
\end{equation}
Thus the resulting force is dominated by the Casimir and  black-body forces, both attractive. For $L=1 \mu{\rm m}$, 
indeed one gets  $\al \simeq 24$. Hence the first internal mode contributes to the thermal
force only with a relative factor
$e^{-24}$! The ratio $\displaystyle \gamma=\frac{\tilde X_{\rm T}} {\tilde
X_{0}}$ is thus, owing to (\ref{Xtot}): $\displaystyle \gamma \simeq - \frac{X_{\rm T}^{\infty}}
{\tilde X_{0}}=\frac 1 3\left(\frac{2\pi}{\al}\right)^4.$
For $L=500\,{\rm nm},\ \al =48$, one gets $\gamma =0.98 \times 10^{-4}$. Thus even at room termperature,
the vacuum fluctuations largely dominate the black-body effects. This is equivalent to a zero-temperature situation,
and sensitive experiments will be able to detect the quantum vacuum effects.

\subsection{{Total free energy}}
Let us finally introduce the {\it total free energy} $F$
associated with the vacuum energy and the
thermal free energy\footnote{In this total free energy the contribution of the exterior volume has been subtracted out.
Its definition coincides with that of the regularized, then renormalized free energy given in sections \ref{sec.Reg}
and \ref{sec.renormalized}
below. One has in particular: $F(0)={\tilde {\cal E}}_{0},\,F(T)-F(0)={\tilde {\cal F}}_{\rm T}$.}
\begin{equation}
\label{Ftot}
F\equiv {\tilde {\cal E}}_{0}
+{\tilde {\cal F}}_{\rm T}=\tilde {\cal E}_{0}
+{\cal F}_{\rm T}- {\cal F}_{\rm T}^{\infty}.
\end{equation}
The total force acting on a plate is therefore
\begin{equation}
X=-\frac{\partial {F}}{\partial L}.
\label{forcetot}
\end{equation}
Using (\ref{energiecasimir}) and (\ref{Ftilde'}), we have
\begin{eqnarray}
\label{G}
F&=&{\cal A}\frac{\pi^2\hbar c}{L^3}
\left[-\frac{1}{720}+{\cal G}(\al)\right],\\
{\cal G}(\al)&=&\frac{1}{\al^3}\left[\sum _{n=0}^{\infty}{}^{\prime}
\psi (\al n)+\frac{2}{\alpha} \zeta(4)\right].
\label{F'}
\end{eqnarray}
The above series yields the natural low-temperature expansion of the thermal function ${\cal G}$ is for $\al \gg 1$.
Owing to definition (\ref{psi})
\begin{eqnarray}
\psi(0)
=-\zeta(3),\,\,\psi(\al)=- (\alpha +1) \left[e^{-\alpha}+\mathcal O\left(e^{-2\al}\right)\right],
\ \al \gg 1,
\end{eqnarray}
whence
\begin{equation}
\label{F''}F
= {\cal A}\frac{{\pi^2}\hbar c} {L^3}\left\{-\frac{1}{720}+\frac{1}{\al^3}\left[
-\frac{1}{2}\zeta(3)+\frac 2 {\alpha} \zeta
(4)- (\alpha +1) \left[e^{-\alpha}+\mathcal O\left(e^{-2\al}\right)\right]\right]\right\}.
\end{equation}
for
$\al=\beta \pi \hbar c /L \gg 1$.\par
The high-temperature expansion can be obtained from (\ref{G})
and the Poisson formula. One can also use a remarkable duality formula
between low and high temperatures
\footnote{This duality is valid only for parallel plates, see section \ref{sec.parallel} below.}\cite{brown}
\begin{eqnarray}
\label{dualG}
\al^2 {\cal G}(\al)={\al'}^2 {\cal G} (\al'),\,\,\al \al'=(2\pi)^2,
\end{eqnarray}
which yields for the thermal free energy:
\begin{equation}
\label{dualite}
{\tilde {\cal F}}_{\rm T}(L,\al)=\left({2 \pi}/{\al }\right)^4{\tilde {\cal F}}_{\rm T}\left(L,
{(2\pi)^2 }/{\al}\right).
\end{equation}
From (\ref{F''}) we deduce the {\it high-temperature} limit of the total free energy:
\begin{equation}
\label{F'''} F
= -{\cal A}
\frac{\zeta(3)}{8 \pi \beta L^2}+ \mathcal O\left(\beta^{-2}e^{-{4\pi^2}/{\al}}\right),\, \al=\beta \pi \hbar c/L \ll 1.
\end{equation}


\section{Force between a sphere and a plane}
\subsection{ Experimental situation}
\begin{figure}
\begin{center}
\includegraphics[angle=0,width=.45\linewidth]{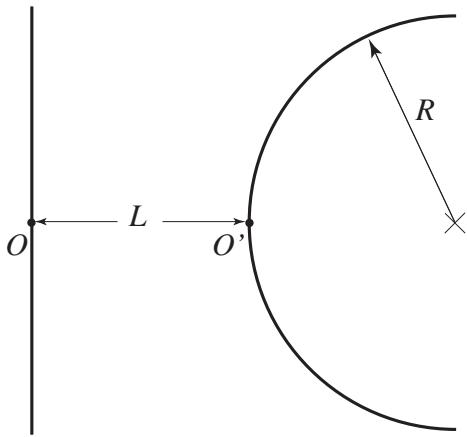}\\
\end{center}
\caption{Experimental set-up of the conducting surfaces.}
\label{platesphere}
\end{figure}

An actual experimental set-up is described in Figure \ref{platesphere}. A metallized sphere of radius $R$
is placed in front of a conducting plate, at a distance $OO'=L$.
In the experiments, like the one performed in 1998 \cite{moh}, a  polystyrene sphere is attached to the arm of an atomic force microscope,
and placed in front of a polished planar surface. Their surfaces are coated with an aluminium layer a few hundred nanometers thick.
To prevent corrosion, they are additionally coated with a very thin alloy layer, which is transparent to the radiation.
The overall radius of the sphere is $R=98.0  \pm 0.25 \
\mu {\rm m}$. The range of distances is $120 \ {\rm
nm}\le L \le 500 \ {\rm nm}$. Measures are performed at room temperature.
Measurements with parallel plates,  
as in the original Casimir's calculation, are also performed \cite{carugno}, but are more 
difficult, due to the necessity to properly align the plates. This experimental problem is avoided 
in the sphere-plate geometry. However,   
the theoretical force is not explicitly known in this case, eventhough it exists in a closed form \cite{BD2} 
(see sections \ref{sec.renormalized} and \ref{two-scatt} below).

\subsection{Derjaguin approximation}
An approximation method, due to Derjaguin
(1934) \cite{der}, allows in the  $L \ll R$ limit, the calculation of the
sph{e}re-plane interaction in terms of the purely planar interaction. One replaces each elementary
slice of the
sphere cut parallel to the plane, by its orthogonal projection towards the plane. The resulting force then is
\begin{equation}
X^{\rm sph}(L)=2\pi R  \int_{L}^{+\infty} \frac 1
{\cal A}{X}(x) \ {\rm d}x,\label{der}
\end{equation}
where $\displaystyle { X}(x)$ is the force (\ref{forcetot}) between two planes at distance $x$.
From (\ref{der}) and definition (\ref{forcetot}) of $X$ we immediately find
\begin{eqnarray}
X^{\rm sph}(L)
=2\pi R \frac{1}{\cal A} F(L),
\label{XF}
\end{eqnarray}
where we used the fact that
the total free energy 
$F$ of two plates vanishes at infinity, as shown by the large-distance or high-temperature equivalent (\ref{F'''}). The Casimir zero-point force on the sphere is thus:
\begin{equation} \label{casimir1}
X_0^{\rm sph}(L)
= 2\pi R \frac{1}{\cal A} \tilde {\cal E}_{0}(L)= - \frac{\pi^3} {360}R\frac{\hbar c}
{L^3},
\end{equation}
where  $\tilde {\cal E}_{0}$ is the Casimir energy (\ref{energiecasimir}) of two plates.
For $R=98.0  \pm 0.25 \,\mu {\rm m},$ and for $L=200\,{\rm
nm}$, one finds for example:
$X_0^{\rm sph}\simeq -33.4\pm 0.09\times 10^{-12}\,{\rm N}.$
Such a force of the order of tens of pico-newtons is perfectly measurable, and comparable to forces
implied in biological systems, e.g., in micromanipulations of single DNA macromolecules.
\begin{figure}[tb]
\begin{center}
\includegraphics[angle=0,width=.9\linewidth]{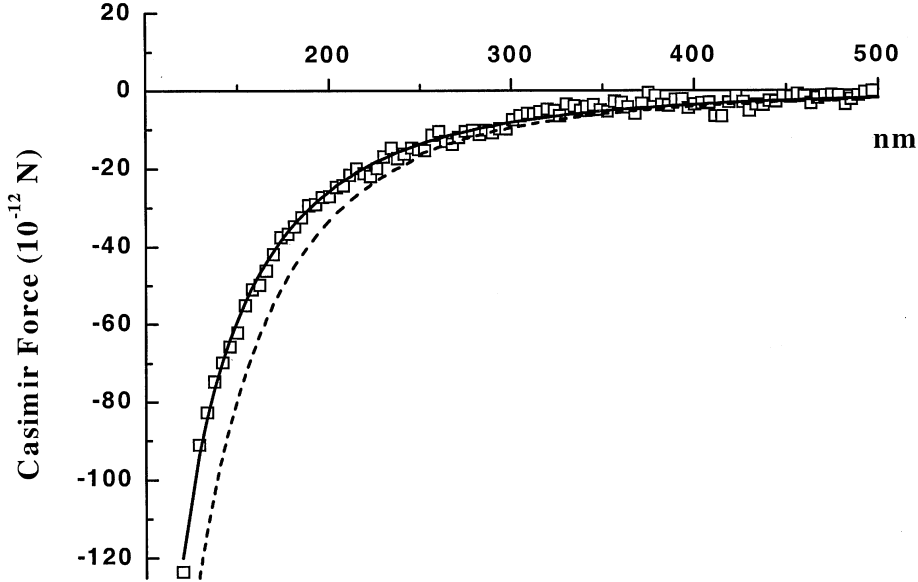}\\
\end{center}
\caption{{Comparison of the zero-point force (\ref{casimir1}) (dashed)
with experimental results; the full curve takes into account finite conductivity and rugosity corrections (ref. \cite{moh}).}}
\label{exp}
\end{figure}
In the experimental range
$120 \,{\rm nm}\le L \le 500 \,{\rm nm}$, the minimal value of  $\alpha$ is
$\alpha_{\rm min}=\alpha (L=500\,{\rm nm})=48$. Hence we can use the short-distance or low-temperature estimate
 (\ref{F''}) of the free energy $F$:
\begin{equation}
\label{casimir4}
X^{\rm sph}(L)= - \frac{\pi^3} {360}R\frac{\hbar c}
{L^3}
\left\{1+{720}\left[
\frac{1}{2{\al^3}}\zeta(3)-\frac 2 {\alpha^4} \zeta
(4)+\mathcal O\left(\alpha^{-2} e^{-\alpha}\right)\right]\right\}.
\end{equation}
The thermal correction in  (\ref{casimir4}) is governed by the
$\zeta(3)$ term, which gives in the experimental range ($L \leq 500\,{\rm
nm}$) a relative thermal correction $4\times 10^{-3}$. It is attractive
and adds to the leading zero-point term (\ref{casimir1}). Formula (\ref{casimir4})
is used by experimenters \cite{moh}. \par


The simple zero-point formula (\ref{casimir1}) is shown in figure \ref{exp}, in comparison with experimental results.
The match is good, eventhough the experimental curve goes further above (\ref{casimir1}) at low values of $L$, with a relative shift in the tens
of percents. This cannot be accounted for by temperature corrections of order $10^{-3}$,
which furthermore lower the theoretical predictions. Good agreement ($\sim 1\%$) is found when finite conductivity and rugosity corrections
are taken into account \cite{moh,harris,reynaud}.
Let us also remark that the analysis of geometrical corrections to the
Derjaguin approximation (\ref{der}), following the general formalism described in the next sections, would be useful.

Before addressing the Casimir effect for arbitrary geometries, let us conclude that one does observe, even at room temperature,
 {\it macroscopic} {e}lectromagn{e}tic forces, generated by vacuum fluctuations, and proportional to
$\hbar c$, and  {\it in the absence of  charge and  photons
in the cavity!} \footnote{It is interesting to note that there is an infinity of longitudinal soft photons
$(n=0, {\bf k} \neq {\bf 0})$ present in the cavity, but they do not contribute to the force;
in contrast,  photons with $n \geq 1$ are essentially absent from the cavity at such short separations $L$,
which allows a direct observation of the vacuum energy, even at room temperature.} Notice that several experiments also give clear evidence of retardation effects in 
 atom-wall interactions, in agreement with the Casimir-Polder prediction \cite{aspect}.

\section{Arbitrary conductor geometries}
\subsection{Eigenmodes}
Let us first consider a single connected region $v$ limited by
perfectly conducting boundaries. The electromagnetic field in $v$ can
be analyzed in terms of the eigenmodes $m$, obtained by solving
Maxwell's equations
\begin{equation}\label{3}
\begin{cases}
{\rm curl}\  {\bf E}_m = {\rm i}\omega_m {\bf B}_m\ ,\ c^2\  {\rm curl}\  {\bf B}_m = -{\rm i}\omega
{\bf E}_m\ ,\\
\text{div}\  {\bf E}_m = 0\ ,\ {\rm div}\  {\bf B}_m = 0,
\end{cases}
\end{equation}
with the boundary conditions (\ref{2}). We shall keep aside the
electrostatic and magnetostatic solutions with zero frequency, which
do not contribute to the Casimir effect. Each mode $m$ behaves as a
harmonic oscillator with frequency $\nu_m = \omega_m/2\pi = c
q_m/2\pi$ where $q_m$ has the dimension of an inverse wavelength. Its associated energy
(\ref{1}) may take the quantized values $\varepsilon(q_m,N) = \hbar
cq_m(N+\frac 1 2)$. At finite temperature it yields a contribution
$f(q_m)$ (\ref{freeenergy}) to the free energy, which can also be written as
(hereafter we use temperature units where $k_B= 1$)
\begin{equation}\label{4}
f(q_m)
= T {\ln} [2\ {\sinh}(\hbar cq_m/2T)]\ .
\end{equation}
At zero temperature the corresponding contribution to the energy of
vacuum is $\frac 1 2 \hbar cq_m$, the limit of (\ref{4}) as $T\to
0$. For the high frequency modes such that $h\nu \gg T$, the free
energy $f(q)$ is dominated by this zero-point energy. For the low
frequency modes such that $h\nu \ll T$,
\begin{equation}\label{5}
f(q) \sim -T \ln(T/\hbar cq)
\end{equation}
is dominated by the classical $-T \ln T$ behavior.

The spectrum of eigenfrequencies, or equivalently of eigenwavenumbers
$q_m^{(v)}$ in the considered region $v$ is
characterized by the {\it density of modes}
\begin{equation}\label{6}
\rho ^{(v)}(q) = \sum_m \delta(q-q^{(v)}_m)\ ,
\end{equation}
and the {\it free energy} of this region is formally equal to
\begin{equation}\label{7}
F^{(v)} = \int_0^{\infty} dq\  \rho ^{(v)}(q) f(q) \ .
\end{equation}

A first difficulty arises if the domain $v$ is infinite, since the
spectrum is then {\it continuous}. We therefore imagine that the full
system is enclosed in a {\it large box} $\Sigma$, with volume $V$, which
will eventually tend to infinity. By assuming this outermost boundary
to be perfectly conducting, we not only discretize the spectrum $m$ in
the open regions $v$, but also confine the field and ensure that no
energy is radiated outwards.

A second difficulty is associated with the fact that the spectra are
{\it not bounded}. Indeed, for large $q$, the distribution (\ref{6})
has the asymptotic expansion
\begin{eqnarray}\label{8}
\rho ^{(v)}(q) &\approx& \frac v {\pi ^{2}} q^2 - \frac 2 {3\pi ^2}\int
\frac{d^2\alpha}{R} + \frac 1 {12\pi ^2} \int ds
\frac{(\pi-\theta)(\pi-5\theta)}{\theta}\\ &+& \mathcal{O}\left(\frac 1{q^2}\right)
+\mathcal{O}(q^{5/2} \times \text{osc})\ .
\nonumber
\end{eqnarray}
The dominant term is proportional to the {\it volume} $v$ of the
considered region; it is the only one which contributes to the
black-body radiation in the thermodynamic limit. The second one is a
{\it curvature} term; it is the integral over the boundaries of $v$
of the average curvature $R^{-1} = \frac 1 2 (R^{-1}_1 + R^{-1}_2)$,
where $R_1$ and $R_2$ are the two main curvature radii at the point
$\alpha$, oriented towards the interior of $v$. It is supplemented, in
case the boundary has not a finite curvature everywhere and includes
wedges with a dihedral angle $\theta$ at the point $s$ of the edge, by
the next, {\it wedge} term, integrated along the wedge; we have for
instance $\theta = \pi/2$ if $v$ is the interior of a cube, $\theta =
3\pi/2$ if it is the exterior. Even if it is smoothed, the
distribution (\ref{8}) finally includes {\it oscillatory} terms with
an amplitude which increases with $q$\footnote{The corresponding semiclassical expansion of the density 
of modes over periodic orbits \cite{BD1} has been recently experimentally tested in a superconducting microvawe cavity 
\cite{demb}}. Since $f(q) \sim \frac 1 2
\hbar cq$ for $q\to \infty$, all the terms exhibited in (\ref{8}) lead
to {\it divergences} in the free energy (\ref{7}).

\subsection{Regularization of the free energy}
\label{sec.Reg}
We encounter here the simplest example of the divergences that plague
quantum field theory. We deal with them by using the standard
technique. We first {\it regularize} the divergent formulae by means
of {\it cut-offs}. We then deduce, from the resulting finite
expressions, quantities that are physically {\it observable} at least
theoretically. We finally {\it renormalize} these quantities by
letting therein the cut-off parameters go to infinity. The theory is
renormalizable if we get a finite limit for the physical quantities.

We have already regularized the ``{\it infrared}'' divergence associated
with the infinite size of vacuum by introducing the {\it box}
$\Sigma$. We shall deal with the ``{\it ultraviolet}'' divergence
associated with the high frequencies $cq\to \infty$ in the integral
(\ref{7}) by introducing, as in section \ref{sec.casimir} above, a {\it cut-off factor} $\chi(q)$ close to 1
for $q \ll Q$ and decreasing sufficiently fast for $q\gg Q$ so as to
restore convergence of the integral. Our final goal is to construct a
renormalized free energy $F$, finite in the limit as $\Sigma \to
\infty$, $Q \to \infty$. If this is feasible, it will mean that the
ideal model of the electromagnetic field outside a set of perfectly
conducting boundaries is renormalizable. In other words the Casimir
effect exists as a property of the field proper, conditioned by the
sole geometry of the boundaries $S$.

Let us first see how one can get rid of the most severe divergence,
associated with the first term of (\ref{8}) which after regularization
with $\chi(q)$ yields $F^{(v)}\propto v \hbar c Q^4$. Consider, for
instance, the Casimir force between two spheres with volumes $v_1$ and
$v_2$. There is here a single empty domain $v_0$, which lies outside
the spheres and inside $\Sigma$. Its volume is $v_0 = V - v_1 -
v_2$. The only universal and natural way to cancel the corresponding
divergence consists in replacing the solid conductors by {\it hollow
  thin conducting shells}, and in taking as a reference the free
energy of the empty enclosure $\Sigma$, which is itself divergent. We now
have three empty domains, $v_0, v_1, v_2$, the volumes of which sum up
to $V$. The most divergent term thus disappears if we substract the
free energy of the empty space within $\Sigma$ from the total free
energy in the presence of the two spherical shells, as
\begin{equation}\label{9}
F^{(v_0)} + F^{(v_1)} + F^{(v_2)} - F^{(\Sigma)}\ .
\end{equation}

\begin{figure}
\begin{center}
\includegraphics[angle=0,width=.5\linewidth]{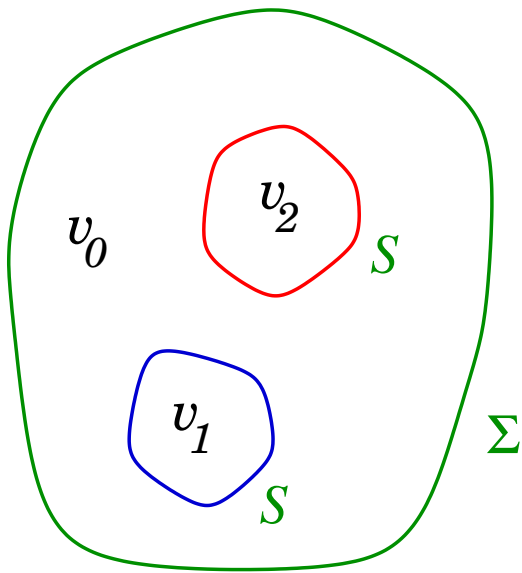}\\
\end{center}
\caption{Set $S$ of surfaces bounding the interiors $v_1$ and $v_2$
of conductors, and defining a third connected (vacuum) domain $v_0$
within enclosure $\Sigma$.}
\label{fi1}
\end{figure}
More generally and more precisely (see Fig. \ref{fi1}), we denote as $S$ the set of
two-dimensional surfaces which bound the considered conductors. They
partition the whole space (within the enclosure $\Sigma$) into a set of
connected regions $v$, some of which coincide with the actual vacuum
(as $v_0$ above), the other ones with the interiors of the conductors
(as $v_1$ and $v_2$ above) . We then define the {\it regularized free
  energy} associated with the {\it whole space partitioned by} $S$ as
\begin{equation}\label{10}
F_{\rm reg} = \int^{\infty}_0 dq \left[\sum_v \rho ^{(v)}(q) - \rho
^{(\Sigma)}(q)\right] f(q) \chi(q) \equiv  \int^{\infty}_0 dq\  \delta\rho(q)
f(q) \chi(q)\ .
\end{equation}
This expression is finite owing to the cut-offs $\Sigma$ and
$\chi(q)$. If, as indicated above, it has a {\it finite limit} $F$ as
$\Sigma\to\infty$ and $\chi(q)\to 1$, independently of the shapes of
$\Sigma$ and $\chi(q)$, the
Casimir effect will appear as a universal property characterized by
the free energy $F$ for the boundaries $S$, which will depend only on
the {\it geometry of} $S$ and on the temperature. This will provide us
with a generating function for all mechanical  and thermal properties
in thermodynamic equilibrium, in two idealized circumstances.

On the one hand, the expression (\ref{10}) can be directly interpreted
as the change in the free energy of the vacuum when a system $S$ of
closed, extremely {\it thin conducting foils} is introduced. The
variations of $F$ under deformations of such foils determine the {\it
  constraints} induced on them by virtual (for $T = 0$) or real
photons (for $T \not= 0$). For instance, for a single sphere
separating two regions $v_1$ and $v_2$, the dependence of (\ref{10}) on the radius
determines the pressure exerted on the skin of this hollow sphere by
the internal and the external field. We shall also encounter below
constraints which tend to corrugate such thin sheets, by studying how
$F$ changes under periodic deformations.

On the other hand, the expression (\ref{10}) is also suited for the
study of forces between {\it bulky indeformable conductors}, as we now
show. Let us return to the above example of two spheres. After
regularization the force between them is associated with the variation
of $F^{(v_0)}$ when they are shifted apart. The regularized quantities
$F^{(v_1)}$ , $F^{(v_2)}$ which enter (\ref{9}) are not physically
relevant to the present problem where we deal with bulky rather than empty
spheres, but they do not depend on the distance between these two
spheres. Actually, a perfectly conducting skin of a sphere behaves as
a perfect screen and the electromagnetic fields, inside and outside, are
independent. Thus the force evaluated from (\ref{9}) is the same as
that evaluated from $F^{(v_0)}$, and it is preferable to use (\ref{9})
because the divergences are expected to be eliminated by this
combination. More generally, whenever solid conducting bodies can be {\it
  displaced but not deformed}, we can derive the forces between them
from the Casimir free energy (\ref{10}) for which the interior of each
body is replaced by vacuum. This trick will allow us to renormalize
$F$. (However, for thermal properties, one should leave aside the
contributions of real photons within the conductors.)

As mentioned above, the cut-off factor $\chi(q)$ which regularizes the integral
(\ref{10}) for large $q$ has a physical meaning. At high frequency,
real conductors are never perfect. Electromagnetic waves can penetrate
them, and go freely across them if they are thin. The objects $S$
become {\it transparent} and the modes within $\Sigma$ tend to be the
same, whether $S$ is present or absent. Thus, for imperfect
conductors, the factor $\delta \rho(q)$ in (\ref{10}) would decrease
for large $q$. In our model we simulate {\it imperfect conduction or transparency} at
high frequency by evaluating $\delta \rho(q)$ for perfectly conducting
sheets $S$ and multiplying by $\chi(q)$.

We now proceed and study the behaviour of (\ref{10}) when the boundary
$\Sigma$ is pushed away to infinity and when the conducting sheets $S$
tend to become perfect with $\chi(q) \to 1$.

\subsection{Fields in the presence of perfect conductors}

Our strategy will rely on the following ideas.
\begin{itemize}
\item[(i)]
We replace the solution of eqs. (\ref{2}), (\ref{3}), which define the
modes in each region $v$, by the determination of the {\it Green
  functions} associated with these partial differential equations and
boundary conditions. Such a Green function contains in a synthetic way
the whole information on the modes. It is a function of a complex
variable $k$, {\it analytic} in the upper half-plane.
\item[(ii)]
We express the distribution of modes $\rho ^{(v)}(q)$ in terms of the
{\it boundary value} for $k\to q + {\rm i}0$ of the Green functions.
\item[(iii)]
This will allow us to regard (\ref{10}) as an integral in the complex
plane $k$ along the half-line $k = q+{\rm i}0$, $q>0$, and to {\it deform
  this contour} towards the pure imaginary axis $k = {\rm i}y$, $y>0$ where
the Green functions are more regular than along the real axis (they have an infinity of poles at $k = \pm q_m$).
\item[(iv)]
We determine the Green functions by means of {\it Neumann's method},
which expresses them as solutions of two-dimensional integral
equations over the boundaries $S$ and $\Sigma$. The regularized free
energy will thereby be expressed through the kernel of these integral
equations in terms of the geometry of $S$ and $\Sigma$.
\item[(v)]
Along the new integration contour $k={\rm i}y$, we can solve these integral
equations by {\it iteration}. The resulting series are convergent, and
can be interpreted physically as describing {\it multiple scattering}
of an electromagnetic wave on the walls, involving {\it successively
  induced currents}.
\item[(vi)]
Convergence of the multiple scattering expansion allows us finally to
control the limit $\Sigma \to \infty$, $\chi \to 1$ and to find an
explicit expression for the limit $F$ of (\ref{10}).
\end{itemize}

We shall content ourselves here with a sketch of this
programme. Detailed proofs can be found in \cite{BD1,BD2}. Given
the symmetry between the fields $\bf E$ and $\bf B$ in eqs.(\ref{3}),
it is convenient to introduce {\it two Green functions}, a {\it
  magnetic} one ${\bf\Gamma}(r,r')$ and an electric one ${\bf
  \Xi}(r,r')$, which are tensors with two indices at $r$ and $r'$. The
first one represents the magnetic field created at the point $r$, in
the presence of the conducting boundaries $S$ and $\Sigma$, by a
magnetic dipole lying at $r'$ and oscillating as $\rm{e}^{-{\rm i}{\it
kct}}$ at the complex frequency $ck/2\pi$. The current density
associated with this source is (within the factor $\mu_0$)
\begin{equation}\label{11}
{\bf j}_0(r,r') = {\rm curl}_r\ [\delta^3(r-r'){\bf 1}]
\end{equation}
where ${\bf 1}$ is the unit tensor. In each region $v,\Gamma(r,r';k)$
is the solution of the partial differential equation
\begin{equation}\label{12}
(\nabla^2 + k^2){\bf\Gamma} = - {\rm curl}\  {\bf j}_0\ \  ,\ \  {\rm div}\  {\bf\Gamma} = 0,
\end{equation}
obtained by eliminating $\bf E$ from the eqs.(\ref{3}), with the
boundary conditions
\begin{equation}\label{13}
{\bf\Gamma}_{\rm n} = 0\ \ ,\ \ ({\rm curl}\  {\bf\Gamma})_{\rm t} = 0.
\end{equation}
The Green function ${\bf\Gamma}(r,r')$ is the magnetic field
generated at the point $r$ by the source ${\bf j}_0$ at $r'$ and the currents
${\bf j}(\alpha,r')$ that it induces at the points $\alpha$ of the
conducting surfaces $S,\Sigma$. We shall denote as $n_{\alpha}$ the
normal vector at $\alpha$, oriented towards the region $v$ where $r$
and $r'$ lie. Both ${\bf j}_0$ and ${\bf j}$ are tensors with two
indices; the first one refers to  the direction of the current, the
second one to the orientation of the dipole at $r'$. Using the
formalism of retarded potentials, we can express the magnetic field
created by each elementary current ${\bf j}$ as $M{\bf j}$, where $M$
is the kernel
 \begin{equation}\label{14}
M(r,r') = {\rm curl}_r\  [G_0(|r-r'|){\bf 1}]\ .
\end{equation}
A product like $M{\bf j}$ stands for integration over space and
summation over a tensor index. The scalar Green function
\begin{equation}\label{15}
G_0(r) = \frac{{\rm e} ^{{\rm i}kr}}{4\pi r}
\end{equation}
is the solution of the equation $(\nabla^2 + k^2) G_0(r) = -\delta
^3(r)$ that vanishes at infinity for ${\rm Im} k > 0$. The magnetic
Green function ${\bf \Gamma}$ is thus expressed as
\begin{equation}\label{16}
{\bf\Gamma}(r,r') = \int d^3r'' M(r,r'')\  {\bf j}_0(r'',r') + \int_{S,\Sigma}
d^2\alpha\  M(r,\alpha)\  {\bf j}(\alpha,r')\ ;
\end{equation}
its first term ${\bf\Gamma}_0$ is the field produced by the dipole
(\ref{11}) in the infinite space. The as yet unkown currents ${\bf j}$
satisfy the integral equation on the boundary
\begin{equation}\label{17}
{\bf j} = {\bf j}_1 + K{\bf j}\ \ ,\ \
{\bf j}_1(\alpha,r') = \int d^3r K(\alpha,r)\  {\bf j}_0(r,r'),
\end{equation}
where the kernel $K$ between two points $\alpha$ and $\beta$ of the
boundary is the tensor
\begin{equation}\label{18}
K(\alpha,\beta) = 2n_{\alpha}\wedge {\rm curl}_{\alpha}[G_0(|\alpha-\beta|){\bf 1}].
\end{equation}
In eq.(\ref{17}) the product $K{\bf j}$ stands for summation on the
tensor index and integration over $\alpha$ on the boundary.
The proof of (\ref{17}),  (\ref{18}) relies on an extension of
Neumann's method, based upon the discontinuity of the surface integral
in (\ref{16}) when $r$ crosses the boundary \cite{BD2}.

Altogether the solution of (\ref{17}) determines ${\bf j}$, and
$\bf\Gamma$ follows from (\ref{16}), taking into account the various
definitions (\ref{11}), (\ref{14}), (\ref{15}), (\ref{18}). The
solution of the partial differential equation (\ref{12}) with the
boundary conditions (\ref{13}) thus amounts to the solution of the
{\it integral equation} (\ref{17}) {\it on the boundary}.

The Green function $\bf\Gamma$ is a {\it generating function for the
  modes} $m$ defined by (\ref{2}), (\ref{3}) in the connected region
  $v$ where the source (\ref{11}) lies. Indeed, in terms of the
  complex variable $k$, its poles are the real points $k = \pm q_m =
  \pm \omega_m/c$ and the corresponding residues are given by
\begin{equation}\label{19}
{\bf\Gamma}(r,r';k) = \sum_m \frac {q^2_m}{q^2_m-k^2}{\bf B}_m(r)
\otimes {\bf B}_m(r')\ ,
\end{equation}
where the magnetic field ${\bf B}_m(r)$ for the mode $m$ is the real
solution of (\ref{2}), (\ref{3}), normalized according to $\int_v d^3r
{\bf B}^2(r) = 1$. The {\it spectral density} (\ref{6}) is thus
related to ${\bf\Gamma}$ through
\begin{equation}\label{20}
\rho ^{(v)}(q) = \frac 2 {\pi q} \int_v d^3 r\  {\rm tr}\  {\rm Im}\
{\bf\Gamma}(r,r;q+i0)\ ,
\end{equation}
where the trace ${\rm tr}$ refers to the tensor indices.

We can likewise introduce an {\it electric Green function} $\bf\Xi$ by
interchanging magnetic and electric fields, which amounts to
interchanging the boundary conditions (\ref{13}). It is the electric
field created at the point $r$ by a source with current density ${\rm
  curl}{\bf j}_0/{\rm i}ck$ in the presence of the boundaries
$S,\Sigma$. Taking (\ref{11}) into account, we see that, except at
$r=r'$, this source produces the same electric field as an electric
dipole with current density $-{\rm i}k[\delta ^3(r-r'){\bf 1}]/c$.

Provided $r$ and $r'$ lie in the same region $v$, the electric Green
function ${\bf \Xi}(r,r';k)$ can be represented by an expression
analogous to (\ref{16}), (\ref{17}), within the mere {\it change in sign
  of} ${\bf j}_1$. (However, whereas $\bf \Gamma$ is expressed by
means of (\ref{16}) in terms of a true electric current density, the
similar expression for ${\bf \Xi}$ involves a fictitious current,
without physical meaning. Moreover, if we take $r$ and $r'$ on
different sides of a boundary $S$, the expression (\ref{16}) for $\bf
\Gamma$ vanishes as it should, whereas the similar representation for
${\bf \Xi}$ provides an unphysical non-zero value. Such a behaviour is
currently found in the books of mathematics that deal with Neumann's
method; in fact, paradoxically, single-layer potentials are used there to
represent Green functions when they vanish on the boundary,
double-layer potentials when their normal derivative vanishes, whereas
it is natural in electrostatics to make the converse choice. It is the
application of Neumann's method in two different ways which allowed us
to find the above simple relation between the representations of $\bf
\Gamma$ and ${\bf \Xi}$.)

Since ${\bf \Xi}$ has the same spectral representation (\ref{19}) as $\bf
\Gamma$ within the replacement of ${\bf B}_m$ by ${\bf E}_m$, we can equivalently express the distribution of eigenmodes (\ref{20}) as
\begin{equation}\label{21}
\rho ^{(v)}(q) = \frac 1 {\pi q} \int_v d^3r\  {\rm tr}\  {\rm Im} [{\bf
\Gamma} + {\bf \Xi}(r,r;q+{\rm i}0)]\ ,
\end{equation}
and simplications will appear owing to this combination.

In fact, the {\it iteration} of the integral equation (\ref{17})
provides
\begin{equation}\label{22}
{\bf j} = {\bf j}_1 + K {\bf j}_1 + K^2 {\bf j}_1 + \cdots \ ,
\end{equation}
a series which exhibits the surface current ${\bf j}$ as the sum of {\it
  successively induced currents}: ${\bf j}_1$ is according to
  (\ref{17}) a current directly induced by the dipolar source on the
  conducting boundaries; it induces in turn through the propagator
  $K$ a secondary current $K{\bf j}_1$, and so on. The expression (\ref{18})
  of $K$ and the behaviour of the free Green function (\ref{15}) show
  that this propagator decreases exponentially at large distances
  (while oscillating) for ${\rm Im} k > 0$. Moreover, $K(\alpha,\beta)$
  vanishes when the point $\beta$ lies in the plane tangent at
  $\alpha$ to the boundary. Thus a current circulating on a plane
  boundary does not induce through $K$ any secondary current on the
  same plane. At short distances $K$ vanishes for a smooth boundary
  and is proportional to its curvature; this ensures the convergence
  of the integrals in $K{\bf j}_1$, $K^2{\bf j}_1$, etc. By relying on these properties, we can show that
  the expansion (\ref{22}) is {\it convergent} at least in the region
  ${\rm Im} k > |{\rm Re} k|$. (For $k=0$ describing static fields, the
  convergence depends on the {\it topology} of the boundaries.) The
  general theory of the Casimir effect will make use of this
  convergence.

The series for $\bf \Gamma$ which results from (\ref{16}) and
(\ref{22}) reads
\begin{equation}\label{23}
{\bf \Gamma} = M \frac 1 {1-K}\  {\bf j}_0 = M\ {\bf j}_0 + MK\ {\bf j}_0 + MK^2\ {\bf j}_0 + MK^3\
{\bf j}_0\ +\ \cdots\ ,
\end{equation}
where $M$ defined by (\ref{14}) describes the propagation of a wave
issued from a unit element of current, and where $K$ defined by
(\ref{18}) describes a similar propagation followed by the creation of
an induced current. We can thus interpret (\ref{23}) as a {\it
  multiple scattering expansion}: the wave issued from the source at
$r'$ may reach directly $r$ (first term); it may propagate from $r'$
to a point $\alpha$ of the boundary where it is scattered to reach $r$
(second term); it may scatter successively twice on the boundary
before reaching $r$ (third term); and so on. The grazing scatterings
vanish, so that the series (\ref{23}) reduces to its first two terms
for a single plane boundary, the second one describing the reflected
wave.

The electric and magnetic Green functions are related to each other
through
\begin{equation}\label{24}
{\bf\Xi}(r,r';k) = k^{-2}\  \text{curl}_r\ {\rm curl}_{r'}\ [{\bf
  \Gamma}(r,r';k) - {\bf \Gamma}(r,r';0)],
\end{equation}
and conversely. However the alternate series
\begin{equation}\label{25}
{\bf\Xi} = M \frac 1 {1+K} \ {\bf j}_0 = M{\bf j}_0 - MK\ {\bf j}_0 +
MK^2{\bf j}_0 -  \cdots\ ,
\end{equation}
which results from the integral equation over $S,\Sigma$ for
${\bf\Xi}$ does not correspond term by term to (\ref{23}), (\ref{24}).

This expansion has the same interpretation in terms of successive scatterings
as (\ref{23}) in the physical situation when $r$ and $r'$ lie in the
same connected region $v$. (However, if $r$ and $r'$ lie in two
different regions $v$ separated by a boundary $S$, the series
(\ref{25}) converges towards some non-zero value, and thus does not
represent the actual electric field produced at $r$ by the source at
$r'$. On the contrary the various terms of the expansion
(\ref{23}) interfere destructively in such a configuration.)

Adding (\ref{23}) and (\ref{25}) as in (\ref{21}) cancels all the odd
terms of the expansion and yields a geometric series in $K^2$. Only
survive in the evaluation of the distribution of modes the terms
describing an {\it even number of scatterings}. On the other hand, in
the summation over the regions $v$, we shall have to transfer the two
points $r$ and $r'$ (with $r=r'$) from one side to the other of the
surface $S$. The kernel $K$ changes its sign in this operation, since
its definition (\ref{18}) involves the normal vector $n_{\alpha}$
oriented in the direction of the domain $v$ where the Green function
is evaluated. After addition of $\bf \Gamma$ and $\bf\Xi$ in
(\ref{21}), the eigenmodes on both sides of $S$ are evaluated with the
{\it same integrand}. This remark will allow us to perform explicitly
the summation over $v$ and the integration over $r$ in (\ref{26}) below.

\subsection{Limiting process}

We are now in position to express the regularized free energy
(\ref{10}) in terms of the kernel $K$ on the boundaries $S,\Sigma$. In
order to switch the integration over $q$ towards the complex plane
$k$, we introduce the {\it generating function of the modes}:
\begin{equation}\label{26}
\delta\Phi(k) = \frac 1 2 \int d^3r\  \text{tr}\ \lim_{r'\to r} \left[\sum_v
({\bf \Gamma}^{(v)} + {\bf \Xi}^{(v)})- {\bf \Gamma}^{(\Sigma)} - {\bf
  \Xi}^{(\Sigma)}(r,r';k)\right ]\ .
\end{equation}
As in (\ref{9}) or (\ref{10}), we have summed over the various regions
$v$ bounded by $S$ and enclosed in $\Sigma$, and substracted the
contribution of the empty box; the integral is therefore carried over
the whole interior of $\Sigma$. The Green functions are singular for
$r'\to r$, but their singular part, which arises only from the first
term ${\bf \Gamma}_0(r-r') = {\bf\Xi}_0(r-r') = M{\bf j}_0$ of ({\ref
  23}) and ({\ref 25}), is
cancelled in (\ref{26}) by the subtraction of ${\bf \Gamma}^{(\Sigma)} + {\bf
  \Xi}^{(\Sigma)}$. Thus no divergence appears in (\ref{26}).

Like the
$\bf\Gamma$ and ${\bf \Xi}$'s, the function $\delta\Phi(k)$ has no other
singularity than poles at the points $k = \pm q_m$ of the real axis,
with residues $\mp \frac 1 2 q_m$. Hence, according to (\ref{21}), we
have
\begin{equation}\label{27}
\delta\rho(q) = \frac 2 {\pi q} {\rm Im}\  \delta \Phi(q+i0)
\end{equation}
for $q>0$. The expression (\ref{4}), (\ref{10}), (\ref{27}) of the
regularized free energy thus reads
\begin{equation}\label{28}
F_{\rm reg} = \frac{2T}{\pi} {\rm Im} \int^{\infty+{\rm i}0}_0 dk \frac{\delta\Phi(k) -
  \delta\Phi({\rm i}0)} k \ln \left(2\, {\sinh} \frac{\hbar ck}{2T}\right) \chi(k)\ .
\end{equation}
The subtraction of the real number $\delta\Phi({\rm i}0)$, which is equal
to $\delta\Phi(0)$ when the box $\Sigma$ is finite, ensures the
convergence of the integral; it corresponds to the fact that static
fields associated with $k=0$ do not contribute to the Casimir effect.

In order to take advantage of the convergence of the expansions
(\ref{23}) and (\ref{25}) for ${\rm Im} k > |{\rm Re} k|$ in (\ref{26}),
(\ref{28}), we shall deform the integration contour in (\ref{28})
towards the imaginary axis. We also {\it get rid of the spatial cutoff}
$\Sigma$. We therefore introduce the function
\begin{equation}\label{29}
\Psi(y) = \lim_{\Sigma\to\infty} \delta\Phi({\rm i}y)\ .
\end{equation}

When taking the limit $\Sigma\to\infty$, we note that $G_0$ defined by
(\ref{15}) and hence $M$ and $K$ defined by  (\ref{14}) and (\ref{18})
decrease exponentially with the distance $l$ as ${\rm e} ^{-yl}$. The
function $\Phi({\rm i}y)$ is represented, through its definition (\ref{26})
and the multiple scattering expansions (\ref{23}), (\ref{25}), by {\it
  closed paths} that bounce an {\it even} number of times on $S$ or
$\Sigma$. All paths which involve only scatterings on $\Sigma$ are
compensated for, owing to the subtraction in (\ref{26}). The
contributions of all the paths which involve at least one back and forth
travel between $S$ and $\Sigma$ contain a factor ${\rm e} ^{-2yL}$ where
$L$ is the minimum distance from $S$ to $\Sigma$; they disappear when
$\Sigma$ is pushed away to infinity. The remaining paths which
contribute to $\Psi(y)$ involve only scatterings on $S$, in even
number; before integration over $r'=r$ of (\ref{26}) they contain for
large $r$ a factor ${\rm e} ^{-2yL}$, where $L$ is the shortest distance
between $r$ and $S$, so that the integral over $r$ is convergent. We
shall perform it explicitly below. Note that the poles on the real
axis of $[\delta\Phi(k) - \delta\Phi({\rm i}0)]/k$ with residues $\pm \frac
1 2$, above which the integral (\ref{28}) runs, become dense in the
limit $\Sigma\to\infty$ and are replaced by a cut.

It remains to {\it get rid of the ultraviolet cutoff} $\chi(k)$ which
eliminates the high frequencies. In order to deform the contour of
(\ref{28}) towards the half line $k={\rm i}y, y>0$ where the singularities
of the logarithm lie, we take for $\chi(k)$ a meromorphic function of
the form
\begin{equation}\label{30}
\chi(k) = \sum_i \frac {a_i}{k^2-\mu ^2_i} + \frac{a ^*_i}{k^2-\mu
  ^{*2}_i}\ .
\end{equation}
The poles $\mu_i$ lie in the first quadrant and their residues satisfy
\begin{equation}\label{31}
\sum_i {\rm Re}\  a_i = 0\ ,\quad \quad -2 \sum_i {\rm Re}(a_i/\mu ^2_i)
= 1\ .
\end{equation}
The moduli $|\mu_i|$ have the same order of magnitude as a number $Q$
that will tend to infinity, so that $\chi(q)$ is close to 1 as long as
$q \ll Q$ and
decreases as $q^{-4}$ when $q \gg Q$. This behaviour ensures the
convergence of (\ref{10}) on account of the behaviour of the residual
terms in (\ref{8}).

Deforming the contour of (\ref{28}) towards the imaginary axis
produces terms associated with the residues at the points $k =
\mu_i$. The properties of $\delta\Phi(k)$ then ensure that these
residues vanish when $|\mu_i|\propto Q \to \infty$.

Indeed, for $k\to\infty$ in the first quadrant, it turns out that the
multiple scattering expansion (\ref{23}), (\ref{25}) is dominated by
its lowest, two-scattering term $MK^2 {\bf j}_0$, provided the surface
$S$ is twice differentiable. The result is found from a short-distance
expansion as
\begin{equation}\label{32}
\Psi(y) = \frac 1 {32\pi} \int_S d^2\alpha \left(\frac 1 {R_1R_2} - \frac
3{R^2}\right) + \mathcal{O}\left(\frac 1 {y^2}\right)\ ,
\end{equation}
where $R_1$ and $R_2$ are the two principal curvature radii and $1/R$
the average curvature at the point $\alpha$. Finally, near the origin,
$\Psi$ behaves as
\begin{equation}\label{33}
\Psi(y) \approx -n(1-Ay) + \mathcal{O}(y^2)\ ,
\end{equation}
where $n$ is the genus of $S$, depending only on its topology ($n=0$ for
a sphere, $n=1$ for a torus), and where $A > 0$.

\subsection{The renormalized casimir free energy}
\label{sec.renormalized}
Altogether, using the above properties, we find the limit of
(\ref{10}) for $\Sigma\to\infty$ and $Q\to\infty$ or $\chi(q)\to 1$ as
\begin{equation}\label{34}
F = \frac{\hbar c}{\pi} \int_0^{\infty} dy[\Psi(y) - \Psi(\infty)] +
2T \int_0^{\infty}\frac{dy}{y} [\Psi(y) - \Psi(+0)]g(y)\ ,
\end{equation}
where the {\it temperature} appears through the sawteeth function
\begin{equation}\label{35}
g(y) = \frac 1 2 - \frac y {\eta} + \sum_{n=1}^{\infty}
\theta(y-n\eta)\ ,\ \eta = \frac{2\pi T}{\hbar c}\ .
\end{equation}
with $\theta(x)=0$ for $x<0$, $\theta(x)=1$ for $x > 0.$
\begin{figure}[tb]
\begin{center}
\includegraphics[angle=0,width=.7\linewidth]{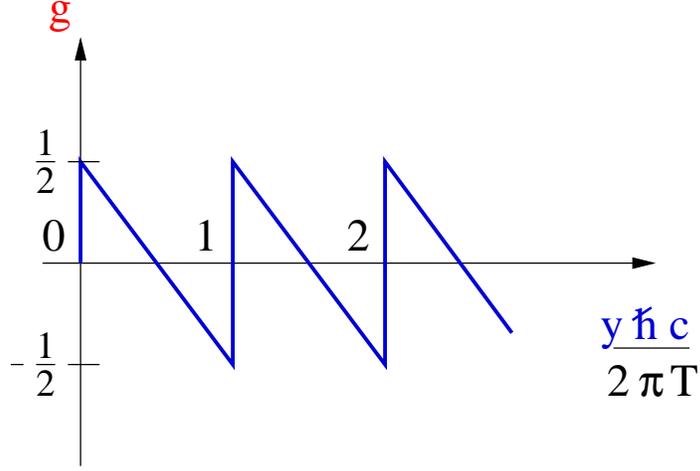}\\
\end{center}
\caption{Sawteeth function $g$ (\ref{35}) drawn as a function of the
dimensionless variable $y/\eta={y\hbar c}/{2\pi T}.$}
\label{fi5}
\end{figure}
A closed expression for the function $\Psi(y)$, which encapsulates the
effect of the {\it geometry} of the boundaries $S$ on
the modes of the field, is found by integrating (\ref{26}) on $r$ in
the whole space. We noted that its integrand is the same for all the
regions $v$ separated by $S$, so that this integral depends only on
the first and last scattering points on $S$ in the expansions
(\ref{23}), (\ref{25}). The integration over $r=r'$ of the
corresponding product of $M$ and $K{\bf j}_0$ can be expressed only in
terms of the kernel $K$ on $S$, and it simply yields $\frac 1 2 y\
dK/dy$.
\begin{figure}[tb]
\begin{center}
\includegraphics[angle=0,width=.65\linewidth]{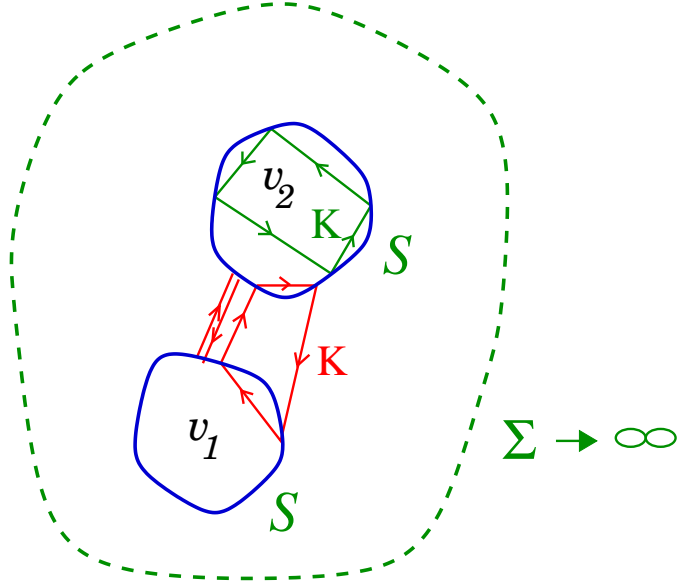}\\
\end{center}
\caption{Closed paths (all with an even number of segments) contributing to the operator trace
${\rm Tr} \ln(1-K^2)$ in
the 
function $\Psi$ (\ref{36}). Multiple scattering on a single shell contributes to the self-energy, while scattering between two 
surfaces contributes to their mutual energy. The enclosure $\Sigma$ (dotted line) has been taken to $\infty$.}
\label{fi2}
\end{figure}
Hence we find
\begin{equation}\label{36}
\Psi(y) = -\frac y 4 \frac d {dy} {\rm Tr} \ln(1-K^2)\ ,
\end{equation}
where the trace ${\rm Tr}$ and the products stand for integration over $S$
of a variable $\alpha$ and
summation on the tensor indices of $K$ (see Fig. \ref{fi2}). Expressed as function of $y$, the kernel
$K$ on the surface $S$ defined by (\ref{15}), (\ref{18}) reads
\begin{equation}\label{37}
K(\alpha,\beta;y) = \frac 1 {2\pi} n_{\alpha} \wedge {\rm
  curl}_{\alpha} \left[\frac{{\rm e}
  ^{-y|\alpha-\beta|}}{|\alpha-\beta|}{\bf 1}\right ]\ .
\end{equation}
It is real, decreases exponentially at large distances, and locally
vanishes as the product of the distance and the curvature for
$|\alpha-\beta|\to 0$. When $S$ consists of several disconnected
pieces, the normals $n_{\alpha}$ on each of them should be oriented
compatibly; for instance, all of them should point towards the
outermost region. Along the
integration path of (\ref{34}), the  series obtained by expanding
$\ln(1-K^2)$ in powers of $K^2$
converges. The values of $\Psi(y)$ at both ends of the integration
path are given by (\ref{32}) and (\ref{33}).

The two terms of (\ref{34}) correspond to the two different phenomena
that we are studying. The first one is the {\it Casimir energy
  proper}, associated with the {\it variation of the zero-point energy} of
the electromagnetic modes that is induced by the introduction of the
perfectly conducting shells. It is the product of $\hbar c$ by a
factor with dimension $L^{-1}$ depending on the shape of $S$. The
second one is the variation of the {\it free energy of the black-body}
due to the effect of the boundaries on the gas of {\it real photons}. It
has no ultraviolet divergence. The temperature $T$ is that of the
walls, which carry random currents in equilibrium with the quantized
field.

When there are several material bodies, the surface $S$ involves
several disconnected sheets. One can then classify the various terms
of the expansion of (\ref{36}) in powers of $K$ according to the
position on these sheets of the scattering points
$\alpha,\beta,\cdots$ (Fig. \ref{fi2}). The terms for which all these points lie on the
same connected sheet describe the free energy of {\it each separate
  body}; they do not contribute to the forces between indeformable
bodies but determine internal constraints for thin conducting
foils. Those for which some propagation $K(\alpha,\beta)$ occurs
between two points $\alpha$ and $\beta$ situated on different sheets
describe the {\it interaction} free energy.

\subsection{Conditions for the existence of the Casimir effect}

We have proved above the existence of a limit $F$ for the free energy
of the field only, for perfectly conducting, thin walls, which defines
the Casimir effect. This has been made possible under two conditions,
which we now discuss.

On the one hand, the Casimir effect exists only owing to the {\it
  electromagnetic} nature of the field. Our proof made use of the
  cancellation of the one-scattering terms $\pm MK{\bf j}_0$ of the
  expansions (\ref{23}) and (\ref{25}), which eliminated a divergent
  surface contribution. Let us show that the {\it
  ultraviolet divergence} of (\ref{10}) {\it cannot be removed for a
  scalar field}, so that the {\it specific features of
  electromagnetism are essential} to assign an energy to the field separately. Actually, the high-frequency
  expansion of the density of eigenmodes that replaces (\ref{8}) for a
  scalar field includes, after a volume term $vq^2/2\pi ^2$, an {\it
  area term} equal to $-sq/8\pi$ for Dirichlet boundary conditions
  (cancellation of the field at the wall), or to $+sq/8\pi$ for
  Neumann conditions (cancellation of the normal derivative), where
  $s$ is the area of the boundary of the domain $v$. This term is
  associated with the occurence of single scattering in the expansion
  analogous to (\ref{23}). In the evaluation of the regularized free
  energy (\ref{10}) the contributions of the two sides of the walls
  add up. A contribution $\mp(\hbar cs/8\pi) \int^{\infty}_0 dq\  q^2
  \chi(q)$ proportional to the area $s$ of the walls $S$ occurs, and
  its divergence when the ultraviolet cutoff $Q$ tends to infinity is
  incurable. A divergent Casimir force, tending either to stretch or
  to shrink the boundaries, would thus appear for a scalar field. The
  properties of matter interacting with the field could not be
  disregarded.

The inexistence of a divergent area term for the electromagnetic
field can be traced to the mixed boundary conditions ${\bf E}_{\rm t} = 0$,
$({\rm curl}{\bf E})_{\rm n} = 0$ and to the constraint ${\rm div}{\bf E} =
0$. The first condition is of the Dirichlet type for the two
tangential components of ${\bf E}$; the constraint yields a Neumann
boundary condition for the normal component. However the condition
$({\rm curl}{\bf E})_{\rm n} = 0$ relates the two tangential components to
each other, so that altogether the electromagnetic field behaves as
two scalar fields, one with Dirichlet, the other with Neumann boundary
conditions. The fact that the area terms are opposite for these two
boundary conditions entails their compensation.

A second condition is also necessary for the renormalization of the
total Casimir energy, namely the {\it smoothness of the boundaries} $S$. We have relied, in
our elimination of the ultraviolet divergence, on the behaviour
(\ref{32}) of $\Psi(y)$ for large $y$, which itself requires a {\it
  finite curvature} of $S$. To understand the origin of this condition, let us return to the expansion (\ref{8}) of the eigenmode
density $\rho^{(v)}$ for a given volume $v$. Its first term was
cancelled by the subtraction $\rho^{(\Sigma)}$ in (\ref{10}). The
next curvature term of (\ref{8}) yields in $F^{(v)}$, given by
(\ref{7}), a $\int q d q$ divergence. Fortunately, at each point
$\alpha$ of the boundary $S$, the curvatures are opposite on the two
sides of this boundary. Hence, in the summation over the domains $v$
of (\ref{10}), the curvature terms cancel one another. Accordingly,
the large $y$ expansion (\ref{32}) of $\Psi(y)$ begins with a second
order curvature contribution.

If, however, the surface $S$ has a {\it sharp fold}, for instance if it is
a hollow thin cube, this cancellation of the divergences from both
sides $S$ no longer occurs. The third, wedge term of (\ref{8}) gives
rise to the same $\int q\ dq$ divergence as the curvature term for each
$F^{(v)}$ . Let us evaluate its coefficient. Going from one side of
the surface $S$ to the other changes the dihedral angle $\theta$ into
$2\pi-\theta$. The sum of the wedge contributions to $\delta\rho$ from
the two neighbouring domains is thus
\begin{equation}\label{38}
\frac 1 {12\pi ^2} \int ds \left
  [\frac{(\pi-\theta)(\pi-5\theta)}{\theta} +
  \frac{(\theta-\pi)(5\theta-9\pi)}{2\pi-\theta}\right ] = \frac 1
  {6\pi} \int ds \frac {(\pi-\theta)^2}{\theta(2\pi-\theta)}\ ,
\end{equation}
and it is a positive number as soon as $\theta \not= \pi$. Hence
$F_{\rm reg}$ is dominated by a positive, divergent term proportional to
(\ref{38}), and the field exerts in this idealized model an infinite
constraint on the considered foil, which tends to {\it flatten} its
dihedron.

The same divergence occurs for an open conducting foil. Its {\it edge}
is equivalent to a dihedral angle $\theta = 2\pi$, the contribution of
which to $\rho ^{(v)}$ is $(3/8\pi) \int ds$. The divergence of $F$
implies, for instance, that the zero-point energy of the field
produces a strong attractive force which tends to join the two halves
of a thin conducting foil that is cut along a line.

We also find a divergence if $S$ involves several adjacent dihedra, for
instance, if it is made of three half-planes joined along their common
edge as in a {\it honeycomb}. In this case, for $\theta =
2\pi/3$, we find a contribution to $\delta\rho$ equal to $(-7/24\pi)\int
ds$ which is {\it negative}. Such a configuration would thus be
particularly stable if the experiment could be realized.

The fact that one cannot define a finite Casimir energy for thin
conducting foils with creases is related to the
singularities of electromagnetic fields near sharp edges.

Nevertheless the above treatment can easily be adapted to Casimir
forces between perfectly conducting {\it rigid bodies}, {\it even if
  they have sharp angles}. Consider, for instance, two bulky
wedges. The free energy of each of them cannot be renormalized,
because the kernel $K(\alpha,\beta)$ is singular as
$1/|\alpha-\beta|^2$ for two neighbouring points $\alpha$ and $\beta$
located on the different sides of the edge; the resulting contribution
${\rm Tr} K^2$ to (\ref{36}), in particular, is seen to generate a
divergence in (\ref{34}). If, however, according to the remark at the
end of section \ref{sec.renormalized}, we focus on the interaction between the two
indeformable wedges, letting aside their own energies, such divergent
contributions are irrelevant. The interaction free energy of the two
wedges is thus obtained by keeping, in the expansion of (\ref{36}) in
powers of $K$, only those terms which involve scatterings on both
wedges (with an even number of factors $K$). Their
contribution to (\ref{34}) is expected to be finite in spite of the
short-distance singularity of $K$ near the edges of the wedges, and
the Casimir force between them is therefore well defined. For example,
for two wedges with the same dihedral angle $2\theta$ facing each
other perpendicularly at a distance $L$, the two-scattering
approximation expressed below by eq.(\ref{50}) provides at $T=0$ an
attractive Casimir interaction energy equal to
$-\hbar c\ {\rm tg}^2\theta/4\pi ^2 L$.

We now review some applications of the general expression (\ref{34})
for the Casimir free energy. More details can be found in \cite{BD2}.

\section{Applications}
\subsection{Parallel plates}
\label{sec.parallel}
For two parallel plates with area $\mathcal A$, lying at a distance $L$ from
each other, the function $\Psi(y)$ associated with the three regions
separated by these plates is
\begin{equation}\label{39}
\Psi(y) = \frac{\mathcal A y^2}{2\pi} \ln(1-{\rm e}^{-2yL})\ ,
\end{equation}
wherefrom we get the elementary Casimir effect at $T=0$. The low
temperature expansion of the free energy (\ref{34}),
\begin{equation}\label{40}
 F(T) = -\frac{\mathcal A\pi^2\hbar c}{720\  L^3} - \frac{\mathcal A T^3\zeta(3) }{2\pi\ \hbar^2c^2}
  + \frac{\mathcal A \pi^2LT^4}{45\hbar^3c^3} + {\mathcal O}(T^2{\rm
 e}^{-\pi \hbar c/LT})\ ,
\end{equation}
agrees with the result (\ref{F''}) of the direct calculation, and shows that the Casimir attraction $-\partial F/\partial L$ increases
with the temperature.

The comparison of (\ref{28}), which includes the factor $\ln(1 - {\rm e}
^{-\hbar c q/T})$, with (\ref{34}), (\ref{39}), exhibits {\it a
  duality between high and low temperatures}, as anticipated in Eqs (\ref{G}), (\ref{dualG}) and (\ref{dualite}),
\begin{equation}\label{41}
F(T) - F(0) = \left(\frac{2LT}{\hbar c}\right)^4 \left[F\left(\frac{\hbar^2 c^2}{4L^2T}\right) -
F(0)\right]\ .
\end{equation}
At high temperature, we have
\begin{equation}\label{42}
F(T) = -\frac{\mathcal A T\ \zeta(3)}{8\pi L^2}  + {\mathcal O}(T^2 {\rm e}^{-4\pi
  LT/\hbar c})\ ,
\end{equation}
yielding again an attraction due to radiation pressure.

The {\it entropy} $-\partial F/\partial T$ rises at low temperatures
as $3\mathcal A  T^2 \zeta(3)/2\pi \hbar^2 c^2$, independently of the distance
between the plates, and tends to a {\it finite} limit $\mathcal A \zeta(3)/8\pi
L^2$ at high temperature.

\subsection{Low temperatures}

Owing to the more and more rapid oscillations of $g(y)$ when $T\to 0$,
we can evaluate the second term of (\ref{34}) for low temperatures by
expanding $\Psi(y)$ around $y=0$. This yields
\begin{equation}\label{43}
F(T) - F(0) = \frac{\pi T^2}{3\hbar c} \Psi'(0) - \frac{\pi^3 T^4}{135
  \hbar^3 c^3} \Psi'''(0) + {\mathcal O}(T^6)\ .
\end{equation}

This behaviour is related to the {\it topology} of the boundaries $S$,
since according to (\ref{33}) $\Psi'(0)$ vanishes for a singly
connected surface, and equals $nA$ for a multiply connected
surface. Accordingly, the low-temperature {\it entropy} arising from
(\ref{43}) behaves for $n=0$ as $T^3$ (like the entropy of the black
body), but is large as $-2\pi\ n\ AT/3\hbar c$ for $n\not= 0$. This
negative sign looks paradoxical. It is related to the fact that for
torus-like topologies, permanent supercurrents can generate {\it
  static magnetic fields}. The occurence of a number $2n$ of such modes
with $q=0$, which do not contribute to the Casimir effect, entails a
depletion in the distribution $\rho ^{(v)}(q)$ for $q\not= 0$. In
fact, we see from (\ref{27}), (\ref{29}), (\ref{33}) that
$\delta\rho(q)$ for $\Sigma \to \infty$ tends to $-2nA/\pi$ as $q\to
0$; it is this negative sign which is reflected in that of the
Casimir entropy. However the total entropy of a quantum system must be
positive. In the present case this results from the positivity of the
total density of eigenmodes $\sum_v \rho^{(v)} = \rho ^{(\Sigma)} +
\delta\rho$ of the field, ensured by the fact that $\rho ^{(\Sigma)}
\sim Vq^2/\pi ^2$ is infinite in the large $\Sigma$ limit considered
here.

The dimensionless parameter of the expansion (\ref{43}) is $\ell T/\hbar
c$ where $\ell$ is the typical size of the system $S$. The lowest order
contributions should become experimentally accessible since this
parameter is 0.5 for $T = 300 {\rm K}$ and $\ell = 3 {\rm \mu m}$.

The different behaviour of (\ref{40}) and (\ref{43}) arises from
the fact that (\ref{43}) holds only for a finite system, whereas for
an infinite system like a pair of parallel plates $\Psi'''(y)$ diverges
when $y\to 0$ as shown by (\ref{39}).

\subsection{High temperatures}

At high temperature the second term of (\ref{34}) is dominated by the
first sawtooth of $g(y)$. The corresponding calculation yields
\begin{equation}\label{44}
F = -{\mathcal C}T \ln(T/\hbar c {\mathcal Q}) + {\mathcal O}(T^{-1})\ ,
\end{equation}
\begin{equation}\label{45}
{\mathcal C} = \Psi(+0) - \Psi(\infty) = \frac 1{32\pi} \int_S
d^2\alpha \left(\frac 3 {R^2} - \frac 1 {R_1R_2}\right) - n\ ,
\end{equation}
\begin{equation}\label{46}
\ln {\mathcal Q} = - \frac 1{{\mathcal C}} \int_0^{\infty} dy \ln y \Psi'(y)\ .
\end{equation}
Here the dimensionless parameter $\hbar c/TR$ of the expansion
(\ref{44}) is governed by a length $R$ of the order ${\mathcal Q}^{-1}$
associated with the short-range behaviour of the kernel $K$. This
characteristic length is therefore a typical curvature radius $R$ of
$S$. The high-temperature limit might become experimentally relevant
for crippled foils with small $R$.

The dominant term of (\ref{44}) is formally the same as the free
energy (\ref{5}) of a number ${\mathcal C}$ of {\it classical harmonic
  oscillators} with average frequency $c {\mathcal Q}/2\pi$. Contrary to what
happens for the black-body which requires a quantum treatment at any
temperature, the Casimir contribution that we have calculated, which
describes the change in the free energy of photons brought in by the
boundaries $S$, takes a {\it classical} form for $T \gg \hbar c/R$.
This is possible here because the modes that contribute to the Casimir
effect have bounded frequencies, whereas the modes with $h\nu \gg T$ crucially
contribute to the black-body radiation.

The internal energy $U \sim {\mathcal C}T$ arising from (\ref{44})
expresses the classical {\it equipartition}, and as usual in the
classical limit the entropy ${\mathcal C}\ln({\rm e}T/\hbar c{\mathcal
  Q})$ depends on
Planck's constant in its additive constant. The number ${\mathcal C}$,
positive or negative, is interpreted as the {\it average number of
  modes} with finite frequency {\it added by the introduction of the
  boundaries} $S$. Here again the {\it topology} of $S$ enters the
expression (\ref{45}) of ${\mathcal C}$ through the genus $n$ and the
integer $\int d^2\alpha/4\pi R_1R_2$. For parallel plates, we have
${\mathcal C}=0$ but ${\mathcal C}\ln Q = -\mathcal A \;\zeta(3)/8\pi L^2$.

The high-temperature Casimir {\it constraints} on the conductors $S$
describe the effects of {\it radiation pressure}. To dominant order
they behave as $T \ln T$ and are obtained by studying how ${\mathcal
  C}$ varies when the conductors are displaced or deformed. They tend
to let $F$ decrease, thus to let ${\mathcal C}$ increase. The only
non-topologic part of ${\mathcal C}$, $3\int d^2\alpha/(32\pi
R^2)$,
does not depend on the relative
position of the conductors. Hence, to the dominant order in $T\ln T$,
there are no forces between different conductors induced by the
field. Moreover, since ${\mathcal C}$ is dimensionless and
scale-invariant, there are no there forces tending to dilate or contract
hollow conducting shells. However ${\mathcal C}$ increases with the
average curvature $1/R$ of $S$, so that the Casimir effect tends at
high temperature to {\it let conducting foils undulate}. This tendency
is limited by the next terms of the expansion (\ref{44}), and the
curvature $|R|^{-1}$ tends to rise up to values of order $T/\hbar c$.

The next order contributions to the constraints, of order $T$, arise
from the contribution ${\mathcal C} T \ln {\mathcal Q}$ to $F$. Since
${\mathcal Q}^{-1}$ is
proportional to the size of $S$, these Casimir forces of order $T$
would tend to contract $S$ if ${\mathcal C}$ is negative, to expand it
if ${\mathcal C}$ is positive.

\subsection{The wrinkling effect}

The existence of constraints that tend to wrinkle conducting surfaces
at high temperature is confirmed by the study of {\it small deformations
  of a thin foil}. The evaluation of (\ref{34}) for a weakly deformed
conducting plane $S$ is achieved by means of a two-dimensional Fourier
analysis. It is found that the constraints created by the field tend
to create ripples with wavelengths larger than $2.9 \hbar c/T$, and to
restore flatness for smaller wavelengths. In particular, the Casimir
effect proper, at zero temperature, tends to suppress the
curvatures. Thus a {\it conducting plane foil} is {\it stable at}
$T=0$, but {\it unstable at} $T\not= 0$ under small deformations.

This phenomena is confirmed by the study of the {\it distribution in
  space of the free energy} of the electromagnetic field. The density
  of Casimir free energy $f(r)$ is obtained in the same way as the
  total free energy (\ref{34}), except for the integration over the
  point  (\ref{26}). In fact, taking into acocunt the spectral
  representation (\ref{19}) of the magnetic Green function
  ${\bf\Gamma}$ and the similar one for the electric function ${\bf
  \Xi}$, the contribution of each pole $k = q_m$ of $\Phi ^{(v)}(k)$
  is then weighted by $\frac 1 2 [{\bf B}^2_m(r) + {\bf
  E}^2_m(r)]$. (The coefficients $\mu ^{-1}_0$ and $\epsilon_0$ of (\ref{1})
  are recovered when the physical fields entering (\ref{3}) are
  expressed in terms of the real and normalized functions entering
  ${\bf \Gamma}$ and ${\bf \Xi}$.) The density of free energy $f(r)$
  is thus found as a series describing a wave starting from $r$,
  scattering an even number of times on $S$ and returning to $r$. The
  two-scattering term is sufficient to provide the free energy density
  at a point $r$ located near $S$, at a distance $d$ from it much
  shorter than the local curvature radii. We find at low temperature
 \begin{equation}\label{47}
f(r) = - \frac{\hbar c}{30\pi^2 R d^3} + \frac{T^3 \zeta(3)}{2\pi R
  \hbar^2c^2} + \cdots\ ,
\end{equation}
and at high temperature
\begin{equation}\label{48}
f(r) = \frac T {16\pi Rd^2} \left[\ln (2dT/\hbar c) + C - \frac 1 4\right]+\cdots\ ,
\end{equation}
where $C$ is Euler's constant.

This density is not bounded when $d\to 0$, and its divergence is
non-integrable. Hence, even after renormalization by subtraction of
the free energy of the vacuum without boundaries, the free energy
$F^{(v)}$ associated with {\it each region} is {\it divergent}. The
total free energy $F = \sum_v F^{(v)}$ is finite because on the two
sides of $S$ the average curvatures $R^{-1}$ are opposite at each
point, so that the divergences from (\ref{47}) or from (\ref{48}) cancel
each other.

The sign in (\ref{47}) shows that the presence of a perfectly
conduting foil produces the transfer of an {\it infinite} amount of
{\it zero-point energy from the concave to the convex side}, whereas,
according to (\ref{48}), the energy of {\it real photons} is
transferred {\it from the convex to the concave} side. These opposite
signs are consistent with the stability or unstability of a plane foil
against deformations, depending on the temperature.

\subsection{Other examples}

\subsubsection{Van der Waals and Casimir-Polder forces.}
The Casimir forces between conductors lying {\it far apart} can be  evaluated
by means of the free energy (\ref{34}) of the vacuum separating
them. We noted that it may also be
attributed to the {\it random currents} that circulate on their
surface and produce the field. Such forces have thus the same nature
as {\it van der Waals forces} of mutual induction, except for the
retarded character of the interaction. We find from
(\ref{34}) that two conductors at a large distance $L$ apart {\it attract}
each other as $1/L^8$ at zero temperature (Casimir-Polder forces \cite{caspol}), as $T/L^7 $ at high
temperature (van der Waals forces). {\it Torques} are also found for anisotropic bodies. The
same results hold for a small conducting body facing a mirror, which is
attracted by it \cite{aspect}.

\subsubsection{Derjaguin approximation.}
\begin{figure}[tb]
\begin{center}
\includegraphics[angle=0,width=.55\linewidth]{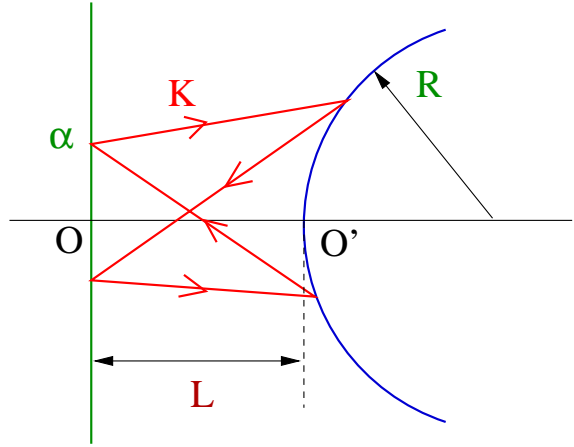}\\
\end{center}
\caption{Multiple scattering contributing to the characteristic function $\Psi$ (\ref{49}).}
\label{fi4}
\end{figure}

For two {\it neighbouring conductors}, the multiple scattering
expansion of (\ref{36}) can be used to justify the {\it Derjaguin
approximation} seen above. Consider for instance the Casimir force between a plane
and a sphere with radius $R$, and denote by $L$ their shortest
distance (Fig. \ref{fi4}).
The trace in (\ref{36}) can be replaced by an integration
over a point $\alpha$ of the plane:
\begin{equation}\label{49}
\Psi(y) = -\frac 1 4 \int d^2\alpha \ y \frac d {dy} \left[{\rm tr}\ln
  (1-K^2)\right]_{\alpha\alpha}\ ,
\end{equation}
where the trace ${\rm tr}$ is meant on the tensor index only. Suppose $L\ll R$. Owing to the exponential decrease of $K$, the
integral (\ref{49}) is dominated by the contributions such that
$\alpha$ lies at a distance $x$ from the sphere of order $L$, and such
that all successive scattering points also lie at a distance of order
$L$ from $\alpha$. Thus, for each $\alpha$, the integrand of
(\ref{49}) is approximately the same as for two parallel plates lying
at a distance $x$ apart, which according to (\ref{39}) is given by
$\pi ^{-1} y^2 \ln(1-e ^{-2yx})$. This is just Derjaguin
approximation. Corrections can be obtained from (\ref{49}).

\subsubsection{Two-scattering approximation.}
\label{two-scatt}
Another useful approximation is the {\it two-scattering
  approximation}, for which we retain for $\Psi(y)$ only the lowest
  order term $\frac 1 2 {\rm Tr} K y dK/dy$ of (\ref{36}).
\begin{figure}[tb]
\begin{center}
\includegraphics[angle=0,width=.55\linewidth]{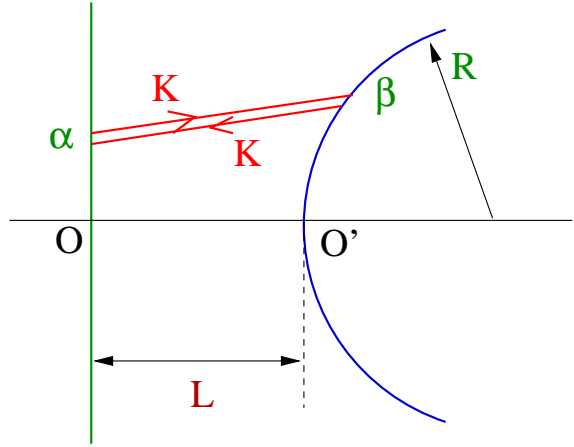}\\
\end{center}
\caption{Geometry of the two-scattering approximation (\ref{50}).}
\label{fi3}
\end{figure}
More explicitly, this yields (Fig. \ref{fi3})
\begin{equation}\label{50}
\begin{split}
\Psi ^{(2)}(y) &= 2y \frac d{dy} \int d^2\alpha d^2\beta
\frac{dG_0(|\alpha-\beta|)} {dn_{\alpha}} \frac{dG_0(|\alpha-\beta|)}
{dn_{\beta}}\\
&= -\frac{y^2}{8\pi ^2}\int d^2{\alpha} d^2{\beta}
(n_{\alpha}.\rho)(n_{\beta}.\rho)\frac 1 {\rho} \frac d{d\rho}
\frac{{\rm e}^{-2y\rho}}{\rho ^2}\\
\end{split}
\end{equation}
where $\rho$ is the vector $\alpha-\beta$. Numerical tests show that
this approximation should be fairly good; for example it yields for
the Casimir force at $T=0$ between two parallel plates the correct
result times $90/\pi ^4$, an error of 8\%.

Using this approximation, we can evaluate the free energy of
interaction between a plane and a sphere. The integrals in (\ref{50})
can be completely worked out and yield
\begin{equation}\label{51}
\Psi ^{(2)}(y) = \left(-\frac 1 2 Ry + \frac 1 4\right) {\rm e}^{-2yL}
- \left(\frac 1 2 Ry + \frac 1 4\right) {\rm e}^{-2yL-4yR}\ .
\end{equation}
Hence we find at $T=0$, for $L \ll R$,
\begin{equation}\label{52}
E ^{(2)} \approx - \frac{\hbar c R}{8\pi L^2} + \frac{\hbar c}{8\pi
  L}\ ,
\end{equation}
which exhibits a correction in $L/R$ to the two-scattering Derjaguin
contribution.

The study of a {\it spherical shell} $S$ with radius $R$ shows that
the Casimir energy at $T=0$ behaves as
\cite{boyer,BD2,kirsten}
\begin{equation}\label{53}
E = 0.046\; \hbar c/R\ .
\end{equation}
At high temperature, we find in agreement with (\ref{44})
\begin{equation}\label{54}
F = -\frac T 4 [\ln(TR/\hbar c) + 0.769] - \left(\frac{\hbar
    c}{R}\right)^2
\frac
1 {3840 T} + {\mathcal O} \left(\frac 1{T^3}\right)\ .
\end{equation}
In the previous examples we had found attractive Casimir forces. In
contrast, these forces tend here at any temperature to {\it expand the
  sphere}. They increase with $T$. The radiation pressure exerted
from inside thus exceeds that exerted from outside, contrary to what
happens for parallel plates.

An intermediate geometry between a sphere and parallel plates is that
of a {\it cylinder}. We evaluate the Casimir energy of a hollow
cylinder $S$ by means of the two-scattering approximation
(\ref{50}). For a cylinder with radius $R$ and length $\mathcal L \gg R$, we
find at low temperature
\begin{equation}\label{55}
F^{(2)} \sim - \frac{2\pi^3}{45\; \hbar^3 c^3} {\mathcal L\; R^2\; T^4}\ ,
\end{equation}
and at high temperature
\begin{equation}\label{56}
F^{(2)} \sim - \frac{3}{64} \frac{\mathcal L\; }{R} T \ln \frac{4.56\; TR}{\hbar c}\ ,
\end{equation}
in agreement with (\ref{44}). The cylinder tends to shrink at high
temperature. At zero temperature, the zero-point energy {\it vanishes}
in the considered approximation and its exact value is very small, an intermediate situation between the
parallel plates (\ref{40}) and the sphere (\ref{53}).

Unfortunately all these Casimir constraints on thin conducting foils,
including the wrinkling effect, are presently difficult to detect
experimentally, because of their weakness compared to the binding
forces that ensure the cohesion of the metallic sheets.


\end{document}